\documentclass{aa}
\usepackage[varg]{txfonts}
\usepackage{graphicx}
\usepackage{natbib}
\begin{document}

\titlerunning{Maunder minimum: A reassessment}
\authorrunning{Usoskin et al.}

\title{The Maunder minimum (1645--1715) was indeed a Grand minimum: A reassessment of multiple datasets}

\author{Ilya G. Usoskin\inst{1,2} \and Rainer Arlt\inst{3} \and Eleanna Asvestari\inst{1} \and Ed Hawkins\inst{6} \and
 Maarit K\"apyl\"a\inst{7} \and Gennady A. Kovaltsov\inst{4} \and Natalie Krivova\inst{5} \and
 Michael Lockwood\inst{6}\and Kalevi Mursula\inst{1} \and Jezebel O'Reilly\inst{6} \and
 Matthew Owens\inst{6} \and Chris J. Scott\inst{6} \and Dmitry D. Sokoloff\inst{8,9} \and Sami K. Solanki\inst{5,10} \and
 Willie Soon\inst{11} \and Jos\'e M. Vaquero\inst{12}}

\institute{
{ReSoLVE Centre of Excellence, University of Oulu, Finland}
\and {Sodankyl\"a Geophysical Observatory, University of Oulu, Finland}
\and {Leibniz Institute for Astrophysics Potsdam, An der Sternwarte 16, 14482 Potsdam, Germany}
\and {Ioffe Physical-Technical Institure, St. Petersburg, Russia}
\and{Max Planck Institute for Solar System Research, Justus-von-Liebig-Weg 3, 37077 G\"ottingen, Germany}
\and{Department of Meteorology, University of Reading, U.K.}
\and{ReSoLVE Centre of Excellence, Department of Computer Science, PO BOX 15400, Aalto University FI-00076 Aalto, Finland}
\and{Moscow State University, Moscow, Russia}
\and {IZMIRAN, Moscow, Russia}
\and{School of Space Research, Kyung Hee University, Yongin, Gyeonggi 446-701, Republic of Korea}
\and{Harvard-Smithsonian Center for Astrophysics, Cambridge, MA, USA}
\and{Departamento de F\'isica, Universidad de Extremadura, M\'erida (Badajoz), Spain}
}

\date{}

\abstract {}
%Aims
{Although the time of the Maunder minimum (1645--1715) is widely known as a period of extremely
 low solar activity, claims are still debated that solar activity during that period might still have been moderate, even higher than the current
 solar cycle \# 24.
We have revisited all the existing pieces of evidence and datasets, both direct and indirect, to
 assess the level of solar activity during the Maunder minimum.}
%Methods
{We discuss the East Asian naked-eye sunspot observations, the telescopic solar observations,
 the fraction of sunspot active days, the latitudinal extent of sunspot positions, auroral sightings
 at high latitudes, cosmogenic radionuclide data as well as solar eclipse observations for that period.
We also consider peculiar features of the Sun (very strong hemispheric asymmetry of sunspot location,
 unusual differential rotation and the lack of the K-corona) that imply a special
 mode of solar activity during the Maunder minimum.
}
%Results
{The level of solar activity during the Maunder minimum is reassessed on the basis of all
 available data sets.
 }
%Conclusions
{We conclude that solar activity was indeed at an exceptionally low level during the Maunder minimum.
Although the exact level is still unclear, it was definitely below that during the Dalton minimum around 1800
 and significantly below that of the current solar cycle \# 24.
Claims of a moderate-to-high level of solar activity during the Maunder minimum
 are rejected at a high confidence level.
}

\maketitle

%_________________________________
\section{Introduction}
\label{S:Intro}

In addition to the dominant 11-year Schwabe cycle, solar activity varies on the
 centennial time scale \citep{hathawayLR}.
It is a common present-day paradigm that the Maunder minimum (MM), occurring
 during the interval 1645--1715 \citep{eddy76}, was a period of
 greatly suppressed solar activity called a Grand minimum.
Grand minima are usually considered as periods of greatly suppressed solar activity
 corresponding to a special state of the solar dynamo \citep{charbonneauLR}.
Of special interest is the so-called core MM (1645--1700) when cyclic
 sunspot activity was hardly visible \citep{vaquero_NA_15}.
While such Grand minima are known, from the indirect evidence provided by the
cosmogenic isotopes $^{14}$C and $^{10}$Be data for the Holocene, to occur
 sporadically, with the Sun spending on average $^1/_6$  of the time in such a
 state \citep{usoskin_AA_07}, the MM is the only Grand minimum covered by direct
 solar (and some relevant terrestrial) observations.
It therefore forms a benchmark for other Grand minima.
We note that other periods of reduced activity during the last centuries, such
 as the Dalton minimum at the turn of the 18th and
 19th centuries, the Gleissberg minimum around 1900, or the weak present solar
 cycle \#24, are also known but they are typically not considered to be Grand minima
 \citep{schussler97,sokoloff04}.
However, the exact level of solar activity in the 17th century remains somewhat
uncertain \citep[e.g.][]{vaquero09,vaquero11,clette14}, leaving
 room for discussions and speculations.
For example, there have been several suggestions that sunspot activity was
 moderate or even high during the core MM (1645--1700),
 being comparable to or even exceeding the current solar cycle \#24
 \citep{schove55,gleissber79,cullen80,nagovitsyn97,ogurtsov03,nagovitsyn04,volobuev04,rek13, zolotova15}.
Some of these were based on a mathematical synthesis using empirical rules in a way
 similar to \citet{schove55} and \citet{nagovitsyn97}, and therefore are not true reconstructions.
Some others used a re-analysis of the direct data series \citep{rek13,zolotova15} and provide
 claimed assessments of the solar variability.
While earlier suggestions have been convincingly rebutted by \citet{eddy83}, the most
 recent ones are still circulating.
If such claims were true, then the MM would not be a Grand minimum.
This would potentially cast doubts upon the existence of any Grand minimum, including those reconstructed from cosmogenic isotopes.

There are indications that the underlying solar magnetic cycles still operated during the MM \citep{beer98,usoskin_JGR_MM_01},
 but at the threshold level as proposed already by \citet{maunder1922}:
\begin{quote}
It ought not to be overlooked that, prolonged as this inactivity of the Sun certainly was, yet few stray spots noted
 during ``the seventy years' death"-- 1660, 1671, 1684, 1695, 1707, 1718 [we are, however, less certain about the exact
 timings of these activity maxima]-- correspond, as nearly as we can expect, to the theoretical dates of maximum. ...
If I may repeat the simile which I used in my paper for {\it Knowledge} in 1894, ``just as in a deeply inundated country,
 the loftiest objects will still raise their heads above the flood, and a spire here, a hill, a tower,
 a tree there, enable one to trace out the configuration of the submerged champaign," to the above mentioned years seem be
 marked out as the crests of a sunken spot-curve.
\end{quote}

The nature of the MM is of much more than purely academic interest.
A recent analysis of cosmogenic isotope data revealed a 10\% chance that Maunder minimum conditions would return within
 50 years of now \citep{lockwood10,solanki11,barnard11}.
It therefore becomes of practical importance to accurately describe and
 understand the MM, since a future Grand minimum is expected to have significant
 implications for space climate and space weather.

Here we present a compilation of observational and historical facts and
 evidence showing that the MM was indeed a Grand minimum
 of solar activity and that the level of solar activity was very low, much lower
 than that during the Dalton minimum as well as the present cycle \# 24 .
In Section~\ref{S:history} we revisit sunspot observations during the MM.
In Section~\ref{S:proxy} we analyze indirect proxy records of solar activity, specifically aurorae borealis and
 cosmogenic isotopes.
In Section~\ref{S:cons} we discuss consequences of the MM for solar dynamo and solar irradiance modelling.
Conclusions are presented in Section~\ref{S:discuss}.

%_________________________________
\section{Sunspot observation in the 17th century}
\label{S:history}

\begin{figure}[t]
\centering \resizebox{\columnwidth}{!}{\includegraphics{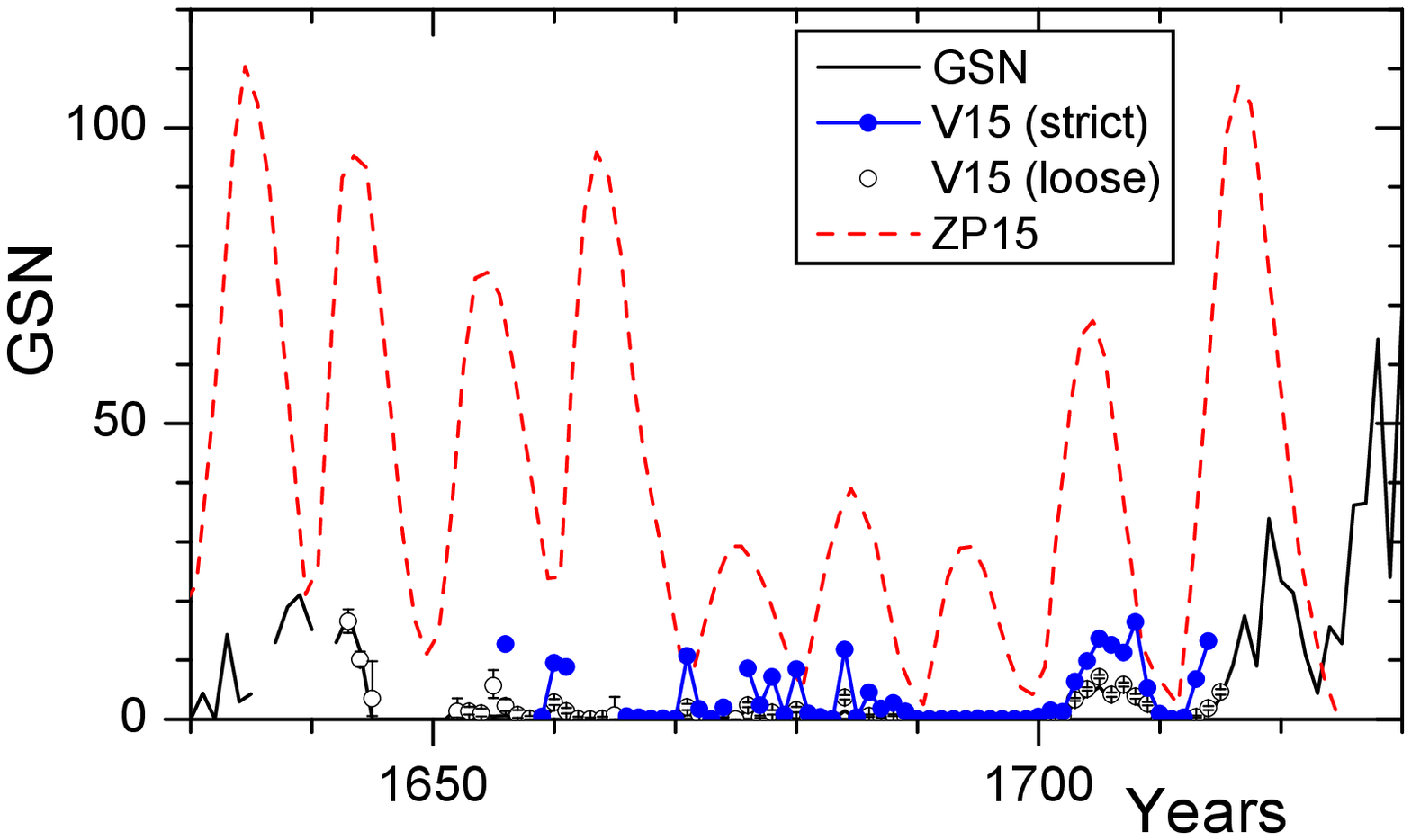}}
\caption{Annual group sunspot numbers during and around the Maunder minimum,
 according to \citet{hoyt98} - GSN, \citet{zolotova15} -- ZP15, and loose and strictly
 conservative models from \citet{vaquero15} (see Sect.~\ref{S:FA}), as denoted in the legend.
}
\label{Fig:SSN}
\end{figure}

Figure~\ref{Fig:SSN} shows different estimates of sunspot activity, quantified
 in terms of the annual group sunspot number (GSN) $R_G$,  around the MM.
The conventional GSN \citep[][called HS98 henceforth]{hoyt98}, with the recent
 corrections, related to newly uncovered data or corrections of earlier errors, applied
  \citep[see details in][]{vaquero11,vaquero_SP_14,lockwood_1_14},
 is shown as the black curve.
This series, however, contains a large number of generic no-spot statements
 (i.e. statements that no spots were seen on the Sun
 during long periods), which should be treated with caution
 \citep[][ see also Sect.~\ref{Sec:gen}]{kovaltsov04,vaquero_rev_07,clette14,zolotova15,vaquero15}.
Figure~\ref{Fig:SSN} shows also two recent estimates of the
 annual GSN by \citet{vaquero15}, who treated generic no-sunspot records
 in the HS98 catalog in a conservative way.
The sunspot numbers were estimated using the active-vs-inactive day statistics
 \citep[see Sect.~\ref{S:FA}, full details in][]{vaquero15}.
All these results lie close to each other and imply very low sunspot
 activity during the MM.
On the contrary, \citet{zolotova15}, called henceforth ZP15,
 argue for higher sunspot activity in the MM
 (the red dotted curve in Figure~\ref{Fig:SSN} is taken from Figure 13 of ZP15),
 with the sunspot cycles being not smaller than a GSN of 30 and even reaching 90--100 during the core MM.

For subsequent analysis we consider two scenarios of solar activity, reflecting
the opposing views on the level of solar activity around the MM before 1749:
(1) L-scenario of low activity during the MM, as based on the conventional
 GSN \citep{hoyt98} with the recent corrections implemented \citep[see][for details]{lockwood_1_14} -- see black curve in Fig.~\ref{Fig:SSN};
(2) H-scenario of high activity during the MM, based on GSN as proposed by ZP15 (red dotted curve in Fig.~\ref{Fig:SSN}).
This scenario qualitatively represents also other suggestions of high activity \citep[e.g.,][]{nagovitsyn97,ogurtsov03,volobuev04}.
After 1749, both scenarios are extended by the International sunspot number
 (http://sidc.oma.be/silso/datafiles).
We use annual values throughout the paper unless another time resolution is explicitly mentioned.

\subsection{Fraction of active days}
\label{S:FA}
High solar cycles imply that $\approx 100$\% of days are active, and sunspots are seen on the Sun almost
 every day during such cycles, except for a few  years around cycle minima \citep{kovaltsov04,vaquero12,vaquero_ASR_14}.
If the sunspot activity was high during the MM, as proposed by the H-scenario, the Sun must have been displaying sunspots almost every day.
However, this clearly contradicts with the data, since the reported sunspot days, also those reported by active observers,
 cover only a small fraction of the year even around the proposed cycle maxima \citep[see Fig.~2 in][]{vaquero15}.
Thus, either one has to assume a severe selection bias for observers reporting only
 a few sunspot days per year while spots were present all the time, or to accept that indeed spots were rare.

During periods of weak solar activity, the fraction of spotless days is a very sensitive indicator of the activity
 level \citep{harvey99,kovaltsov04,vaquero_SP_14}, much more precise than the sunspot counts.
However, this quantity tends to zero (almost all days are active) when the average sunspot number
  exceeds 20 \citep{vaquero15}.
\citet{vaquero15} considered several statistically conservative models to assess the sunspot number during the MM
 from the active day fraction.
The "loose" model ignores all generic no-spot statements and accepts only explicit no-spot records with exact date
 and explicit statements of no spots on the Sun, while the "strict" model considers only such explicit statements as
 in the "loose" model but made by at least two independent observers for the spotless days.
In this way, the possibility of omitting spots is greatly reduced since the two observers would have to omit the
 same spot independently.
The strict model can be considered as the most generous upper bound to sunspot activity during the MM.
However, it most likely exaggerates the activity by over-suppressing records reporting no spots on the Sun.
These models are shown in Fig.~\ref{Fig:SSN}
One can see that these estimates yield sunspot numbers that do not exceed 5 (15) for the "loose" ("strict") model
 during the MM.

\subsection{Occidental telescopic sunspot observations: Historical perspective}
\label{S:telescop}
The use of the telescope for astronomical observations became widespread quickly after 1609.
We know that there were telescopes with sufficient quality and size to see even small spots,
 in the second half of the 17th century.
It is also known that astronomers of that era used other devices in their routine observations such as mural quadrants
 or meridian lines \citep{heilborn99}.
However, as proposed by ZP15, the quality of the sunspot data for that period might be compromised by non-scientific biases.

\subsubsection{Dominant world view}
\label{Sec:ww}
Recently, ZP15 suggested that scientists of the 17th century might be influenced by the ``dominant
 worldview of the seventeenth century that spots (Sun's planets) are shadows
 from a transit of unknown celestial bodies'', and that ``an object on the solar surface with an
 irregular shape or consisting of a set of small spots could have been omitted in a textual report because
 it was impossible to recognize that this object is a celestial body''.
This would suggest that professional astronomers of the 17th century, even if technically capable of observing spots,
  might distort the actual records for politically/religiously motivated, non-scientific reasons.
This was the key argument for ZP15 to propose the high solar activity during the MM.
Below we discuss that, on the contrary, scientists of the 17th century were reporting sunspots quite objectively.

{\it Sunspots: Planets or solar features?}

\noindent
There was a controversy in the first decades of the 17th century about location of sunspots: either on the Sun (like clouds),
 or orbiting at a distance (as a planet).
However, already Scheiner and Hevelius plotted non-circular plots and showed the perspective foreshortening
 of spots near the limb.
In his Accuratior Disquisito, Christoph \citet{scheiner1612} wrote pseudonymously as `Appelles waiting behind the picture'
 and detailed the appearance of spots as of irregular shape and variable, and finally concluded \citep{galileo10}:
\begin{quote}
They are not to be admitted among the number of stars, because they are of an irregular shape, because they change their shape,
 because they [\dots] should already have returned several times, contrary to what has happened,
 because spots frequently arise in the middle of the Sun that at ingress escaped sharp eyes,
 because sometimes some disappear before having finished their course.
\end{quote}
Even though Scheiner had believed until this point that sunspots were bodies or other entities just outside the
 Sun, he did note all their properties very objectively.
Later, \cite{scheiner1630} concluded in his comprehensive book on sunspots, ``Macul\ae{} non sunt extra solem''
(spots are not outside the Sun, p.~455ff.) and even ``Nuclei Macularum sunt profundi'' (the cores of sunspots
are deep, p.~506).
On the contrary, \citet{smogulecz_schoenberger1626} who were colleagues of Scheiner in Ingolstadt and Freiburg-im-Breisgau,
 respectively, called the spots ``stella\ae{} solares'' (=~solar stars) which was meant in the sense of moons.
Some authors, especially anti-Copernican astronomers, such as Antonius Maria Schyrleus
 of Rheita (1604--1660) \citep[see][]{gomez15} and Charles Malapert (1581--1630), followed the planetary model.
On the other hand, Galileo had geometrically demonstrated (using the measured apparent
 velocities of crossing the solar disc) that spots are located on the solar surface.
In fact, the changes in the trajectory of sunspots on the "solar surface" were an important element of discussion
 in the context of heliocentrism \citep{smith85, hutchison90, topper99}.

It was clear already in that time that sunspots are not planets, for reasons of
 the form, color, shape of the spots near the limb and their occasional disappearance in the middle of the disk.
A nice example of the kind is given in a letter to William Gascoigne (1612--1644),
 that William Crabtree wrote on 7 August 1640 (= Aug 17 greg.) \citep{chapman04}, as published by \cite{derham1711}:
\begin{quote}
I have often observed these Spots; yet from all my Observations
cannot find one Argument to prove them other than fading Bodies.
But that they are no Stars, but unconstant (in regard of their
Generation) and irregular Excrescences arising out of, or
proceeding from the Sun's Body, many things seem to me to make
it more than probable.
\end{quote}

Although some astronomers still believed in the mid-17th century that sunspots were small
 planets orbiting the Sun, the common paradigm among the astronomers of that time was ``that spots were current
 material features on the very surface of the Sun'' \citep[][page 78]{brody02}.
Therefore, observers of sunspots during the MM, in particular professional astronomers, did not adhere to
the "dominant worldview" of the planetary nature of sunspots and hence were not strongly influenced
 by it, contrary to the claim of ZP15.

{\it Galileo's trial.}

\noindent
We note that the problem in the trial of Galileo was not the Copernican system, but the claim that astronomical
 hypotheses can be validated or invalidated (an absurd presumption for many people of the early 17th century) and the potential
 claim of re-interpreting the Bible \citep{schroeder2002}.
The planetary system was considered as a mathematical tool to compute the motion of the planets as precisely as possible;
 it was not a subject to be proven.
This subtle difference was an important issue in the first half of the 17th century to comply with the requirements of
 the catholic church.
While an entire discussion of the various misconceptions about the Galileo trial is beyond the scope of this Paper,
 there are many indications that the nature and origin of celestial phenomena had been discussed by the scholars of
 the 17th century, rather than discounted by a standard world view.
We are not aware of any documental evidence that writing about sunspots was prohibited or generally disliked by the majority
 of observers in any document.

{\it Shape of sunspots.}

\begin{figure}[t]
\centering \resizebox{\columnwidth}{!}{\includegraphics{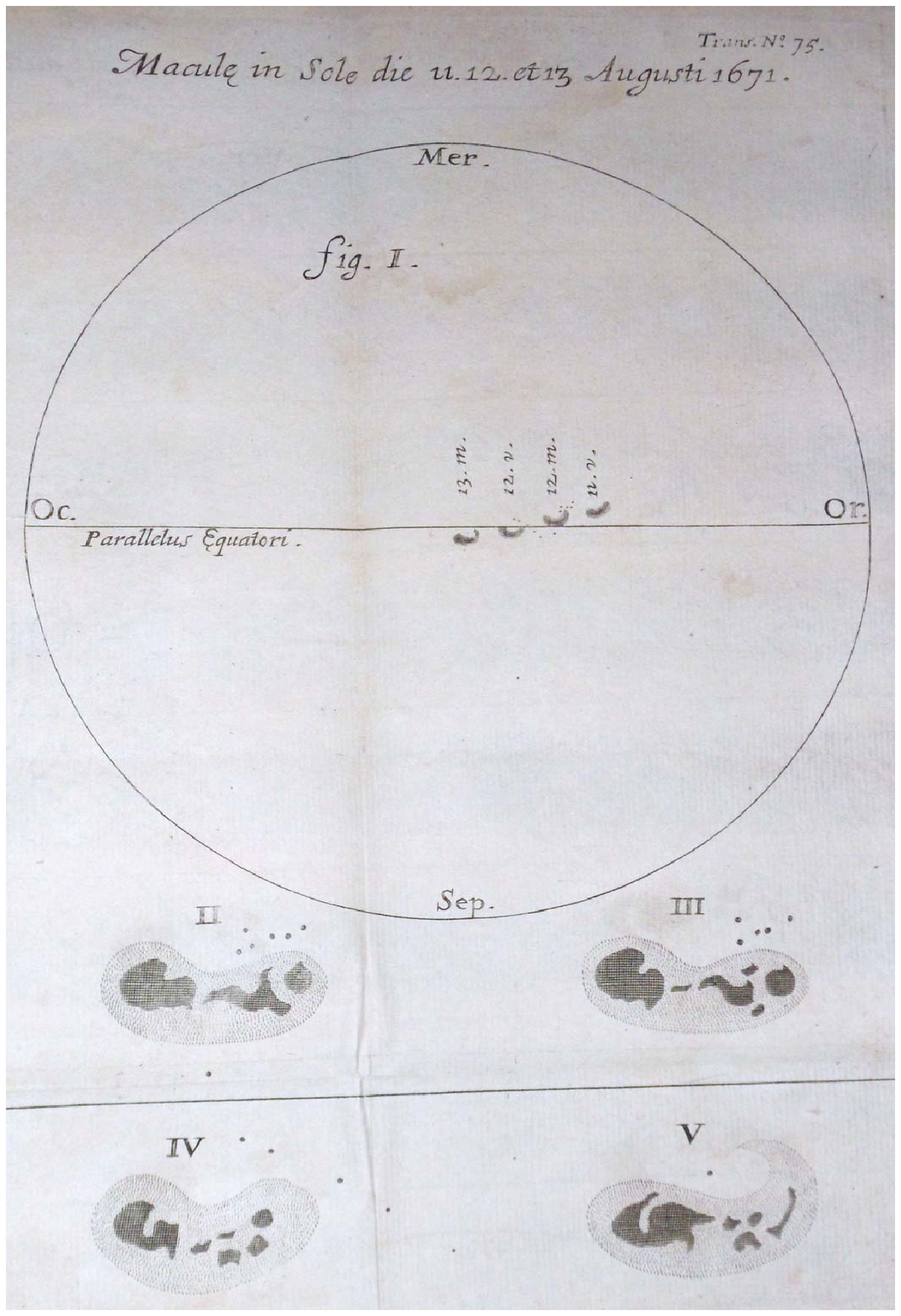}}
\caption{Drawing of a sunspot group observed in August 1671, as published in number 75 of the Philosophical Transactions,
 corresponding to August 14, 1671.
}
\label{Fig:cassini2}
\end{figure}

\begin{figure}[t]
\centering \resizebox{\columnwidth}{!}{\includegraphics{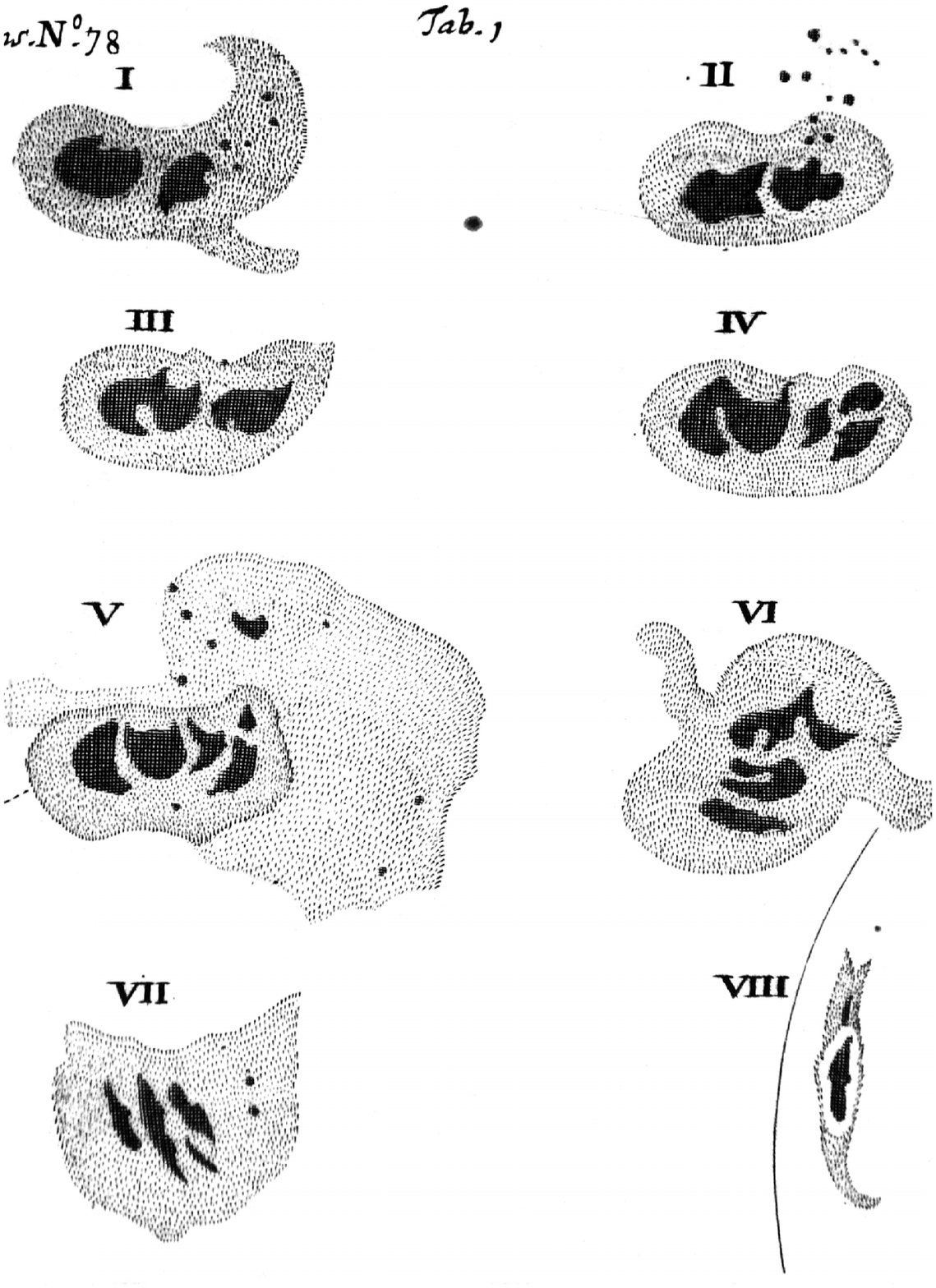}}
\caption{Sunspot drawing of by G.D. Cassini in 1671 \citep{oldenburg1671_3}.
}
\label{Fig:cassini3}
\end{figure}

\begin{figure}[t]
\centering \resizebox{\columnwidth}{!}{\includegraphics{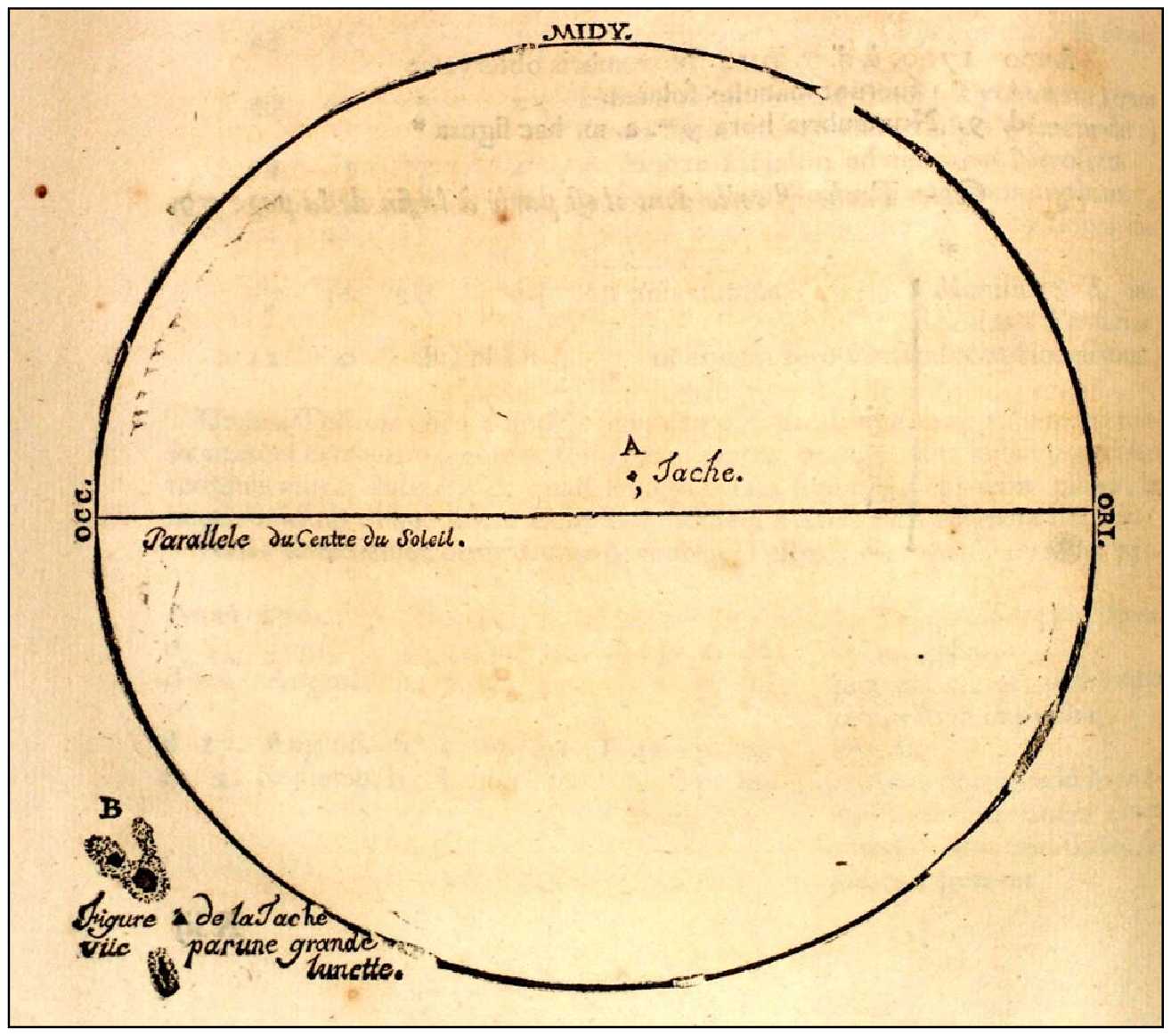}}
\caption{Sunspot observed by J. Cassini and Maraldi from Montpellier (Mar-29-1701).
Reproduced from page 78 of the Histoire de L'Acad\'emie Royale des Sciences (Ann\'ee MDCCI).
}
\label{Fig:cassini}
\end{figure}

\noindent
ZP15 presented a hand-picked selection of a few drawings to support their statement that
 ``there was a tendency to draw sunspots as objects of a circularized form'', but there are plenty of other
 drawings from the same time showing sunspots of irregular shape and sunspot groups with complex structures.
Here we show only a few of numerous examples.
Figure~\ref{Fig:cassini2} depicts a sunspot group observed in several observatories in Europe in August 1671.
A dominant spot with a complex structure having multiple umbrae within the same penumbra and a group of small spots in
its surroundings can be appreciated.
Another example (Figure~\ref{Fig:cassini3}) shows a spot drawing by G.D. Cassini in 1671 \citep{oldenburg1671_3}
 \footnote{Henry Oldenburg was Secretary of the Royal Society and compiled findings from letters of other scientists in the Philosophical
 Transactions in his own words. We therefore cite his name although it is not given for the actual article.},
 which illustrates the complexity and non-circularity as well as the foreshortening of sunspots very clearly.
Finally, Figure~\ref{Fig:cassini} displays a sunspot observation made by J. Cassini and Maraldi from Montpellier (Mar-29-1701).
There is a small sunspot group (labelled as A) approximately in the middle of the solar disc
 which is zoomed in the bottom left corner, exhibiting a complex structure, and a legend that reads ``Shape of the Spot observed
 with a large telescope''.
These drawings are not limited to ``circularized forms'', and such instances are numerous.

It is important that observers who made drawings actually retained the perspective foreshortening of the spots near the solar limb.
\citet{galileo1613}, \citet{scheiner1630}, \citet{hevelius1647}, G.D. Cassini in 1671 \citep{oldenburg1671_3} as well as
 \citet[observation of 1684]{cassini1730}, P.~\citet[observation of 1703]{delahire1720}, and \citet{derham1703}
 all drew slim, non-circular spots near the edge of the Sun.
It was clear to them that those objects cannot be spheres.
They were not shadows either since that would require an additional light source similar to the Sun
 which is not observed.
A note by G.D. Cassini of 1684 says \citep{cassini1730}:
\begin{quote}
This penumbra is getting rounder when the spot approaches the center,
as it is always happening, this is an indication that this penumbra
is flat, and that it looks narrow only because it is presenting
itself in an oblique manner, as is the surface of the Sun towards
the limb, on which it has to lie.
\end{quote}
While G.D. Cassini was an opponent of Copernicus and Newton \citep{habashi07}
 and in fact discovered a number of satellites of Saturn, he did accept that
 sunspots are on the solar surface and did not alter their appearance to
 make them circular.

Thus, the idea suggested by ZP15 of the strong influence of theological or philosophical ideas
 about the perfection of the celestial bodies (especially the Sun) on professional astronomers in the
 late 17th century is not supported by our actual knowledge of solar observations and scientific vision during that time.
There is sufficient evidence that, beginning with the use of telescopes in astronomy, the existence and the
 nature of sunspots were thoroughly discussed, including various opinions, but based on the best observation
 technology of the time.
We further conclude that sunspots were not omitted deliberately from observing records for
 religious, philosophical or political reasons during the MM.
The observational coverage was just incomplete and partially vague.
Moreover, many existing pieces of evidence imply that spots of different shapes were recorded, contrary to
 the claim of ZP15.

\subsubsection{The very low activity during 1660--1671}
\label{S:1660}
The period of 1660--1671 is indicative of very low activity in the HS98 database, but it is mostly based
 on generic statements of the absence of sunspots.
For example, based on a report by G.D.~Cassini, a sunspot observed in 1671 \citep{oldenburg1671_2}
 was described in detail, and it was noted that
\begin{quote}
 ``it is now about twenty years since, that
 Astronomers have not seen any considerable spots in the Sun, though before that time [\dots] they have from time
 to time observed them.
The Sun appeared all that while with an entire brightness.''
\end{quote}
The last sentence implies that the Sun was also void of any other dark features, even if they would not have been
 reported in terms of sunspots.
There is also a footnote saying that indeed some spots were witnessed in 1660 and 1661, so the 20~years
mentioned are exaggerated.
The Journal also states \citep{oldenburg1671_1} that ``as far as we can learn, the last observation in England
 of any Solar Spots, was made by our Noble Philosopher Mr. Boyl'' on Apr~27 (=~May~7 greg.) 1660 and May~25
 (=~Jun~4 greg.).
He described a ``very dark spot almost of quadrangular form''.
Additionally, one of the spots was described as oval, while another was oblong and curved.
This statement contradicts the assumption by ZP15 that the majority of non-circular spots were omitted, especially in
 conjunction with the surprise with which the article was written that spots were seen at all.
If there were a number of non-circular spots during this 10-year period (allegedly not reported), there would have been no
 reason to `celebrate' yet another non-circular spot in 1671.

As another example, \cite{spoerer1889}, p.~315, cited a note by Weigel from Jena in 1665 which can be translated as
\begin{quote}
Many diligent observers of the skies have wondered here that for such a long time no
 spots were noticeable on the Sun.
And we need to admit here in Jena that, despite having tried in many ways, setting up
 large and small spotting scopes pointed to the Sun, we have not found such phenomena
 for a considerable amount of time.
\end{quote}
Since the notes on the absence of spots come from various countries and from catholic, protestant and Anglican people,
 we believe there was no wide-spread religious attitude to withhold spots in order to save the purity of the Sun.

The only positive sunspot report between 1660 and 1671 in the HS98 database is the one by Kircher in 1667.
This data point comes from a note \citep[p.~49]{frick1681} stating that
\begin{quote}
the late Christoff Weickman, who was experienced in optics
 and made a number of excellent telescopes, watched the Sun at
 various times hoping to see the like [sunspots] on the Sun,
 but could never get a glimpse of them [\dots]
So Mr Weickman wrote to Father Kircher and uncovered him that
 he could not see such things on the Sun, does not know why
 this is or where the mistake could be.
Father Kircher answered from Rome on 2~September 1667 that it happens very rarely
 that one could see the Sun as such; he had not seen it in such
 a manner more than once, namely Anno 1636.
\end{quote}
One can see that the date of the letter in 1667 was mistakenly considered as the observing date.
Instead the report clearly indicates that no sunspots were seen at all by Weickman in the 1660s.
The sunspot observation by Kircher in 1667 is erroneous and needs to be removed from the HS98 database.
Then no indication of sunspots exists in the 1660s.
We note that this false report was used by ZP15 to evaluate the sunspot cycle maximum around that date.

\subsection{Generic statements and gaps in the HS98 database}
\label{Sec:gen}

The database of HS98 forms a basis to many studies of sunspots records during the period under investigation.
In particular, ZP15 based their arguments on this database without referring to the original records.
However, it contains a number of not obvious features which can be easily misinterpreted if not considered properly.
Here we discuss such features which are directly related to the evaluation of sunspot activity in the 17th century.

In particular, many no-spot records were related to astrometric observations of the Sun
 such as solar meridian altitude or the apparent solar diameter \citep{vaquero_ASR_14a}.
For example, \citet{manfredi1736} listed more than 4200 solar meridian observations made by several scientists during
 1655-–1736 using the gigantic camera obscura installed on the floor of the Basilica of San Petronio in Bologna.
Such observations were not focused on sunspots and did not include any mentioning of spots.
However, HS98 treated all these reports as observations of the absence of sunspot groups which, of course, was incorrect.

HS98 database contains gaps of the observing records of Marius and Riccioli, that occur exactly
 during days when other observers reported spots, which was interpreted by ZP15 as indications that they
 deliberately stopped reporting to hide sunspots: ``It is noteworthy that when the Sun became active,
 Marius and Riccioli immediately stopped observations.''.
We note that this interpretation is erroneous and based on ignorance of the detail of the HS98 database
 as explained below.

The original statement by Marius from Apr~16 (=~Apr~26 greg.) 1619, on which this series is based, is
%\begin{quote}
%\dots dieweil ich nun \"uber die anderhalb Jahr nicht mehr so viel maculas in disco Solis hab finden k\"onnen/
%ja gar offt kein einig maculam antroffen/ das doch vorige Jahr niemals geschehen/
%dahero ich dann in meinen observationibus verzeichnet/
%Mirum mihi videtur, adeo raras vel s\ae{}pius nullas maculas in disco solis deprehendi,
%quod ante h\^ac nunque est observatum.
%\end{quote}
\begin{quote}
While I did not find as many spots in the disk of the Sun over the past one-and-a-half years, often
 not even a single spot, which was never seen in the year before, I noted in my observing diary:
Mirum mihi videtur, adeo raras vel s\ae{}pius nullas maculas in disco solis deprehendi,
 quod ante h\^ac nunque est observatum
\end{quote}
which is a Latin repetition of what he said before.
Marius clearly states that the sunspot number was not exactly zero, but very low.
HS98 have used this statement to approximate the activity by zeros in their database,
 more precisely by filling all dates of the 1.5-yr interval with zeros except the periods
 when other observers did see spots.
The existence of the gaps is by no means based on the actual observing report by Marius, but is an
 effect of the way HS98 have interpreted the comment.

The same reason holds for the gaps in the sunspots reported by \citet{riccioli1653}, p.~96, whose data (zeros)
 in the HS98 database are based on the statement that
\begin{quote}
\dots in the year 1618 when a comet and tail shone, no spots were observed, said
 Argolus in Pandosion Sph\ae{}ricum chapter 44.
\end{quote}
The original statement by \cite{argolus1644}, p.~213, states:
``Anno 1618 tempore quo Trabs, et Cometa affulsit nulla visa est.''
Apart from the fact that it was not Riccioli himself who observed, this again led to filling all days in 1618
 with zeros (in the HS98 database) except the days when other observers saw spots.

The method of filling the HS98 database for many months and even years with zeros is based on generic verbal reports on
 the absence of spots for long periods also in the cases of Picard, G.D. Cassini, Dechales, Maraldi, Siverus and others
 \citep[see, e.g,][]{vaquero11,vaquero15}.
HS98 must have filled those periods in the sense of probably very low activity, but they are not meant to provide exact
 timings of observations, as ZP15 interpreted them.
The appearance of gaps in zero records when other observers reported spots is not an indication of withholding spots
 in observing reports but rather a simple technical way of avoiding conflicting data in the HS98 database.
ZP15 mistook the entries in the HS98 database for actual observing dates and interpreted them incorrectly.

While assuming a large number of days without spots is a significant underestimation of the solar activity on the
 one hand, as shown by \cite{vaquero15} and has been pointed out by ZP15 as well, the assumption that
 observers deliberately stopped reporting is, on the other hand, not supported by any original text and
 remains an ungrounded speculation.

The observations by Hevelius of 1653--1684 as recovered by \cite{hoyt95} should also be scrutinized
 with regard to a possible omission of spots.
Citing the former reference, ZP15 even claim that ``Hevelius quite consciously did not record sunspots'', while the
 original statement was that ``Hevelius occasionally missed sunspots but usually was a reliable observer.''
Actually, out of 24~groups that could have been detected by Hevelius given his observing days, he saw 20 \citep{hoyt95}.
He never reported the absence of sunspots when others saw them.
The four occasions are simply not accompanied by any statement about presence or absence of spots.
This can be understood considering that the sunspot notes are just remarks on his solar elevation measurements
 \citep[part 3]{hevelius1679}.
Those, however, were made with a quadrans azimutalis which had no telescope, since Hevelius refused to switch to
 a telescope at some point, perhaps because he did not want to spoil his time series of measurements \citep{habashi07}.
He therefore could not see sunspots at all with his device and had to take an additional instrument to observe them,
 and it is probable that he did not do so on each day
 he measured the solar elevation, hence left so many days neither with positive nor with negative information on
 sunspots.
We have to treat those as non-observations.

\subsection{Methodological errors of ZP15}
\label{S:error}
The original work by ZP15 unfortunately contains a number of methodological errors
 which eventually led them to an extreme conclusion that sunspot activity during the MM was at a moderate to high level.
In particular, ZP15 sometimes incorrectly interpreted published records.
Moreover they used the original uncorrected record of HS98,
 while numerous corrections have been made to that during the last 17 years \citep[e.g.][]{vaquero11,vaquero_SP_14,carrasco15}.
Here we discuss some of the errors in ZP15, as examples of erroneous interpretation of historical data, in detail.

\subsubsection{Sunspot drawings vs. textual notes}
According to ZP15 ``sunspot drawings provide a significantly larger number of sunspots, compared to textual or tabular sources''.
This is trivial considering that the tabular sources often are related to astrometric observations of the Sun such as solar
 meridian altitude or the apparent solar diameter \citep{vaquero_ASR_14a}.
However, if one considers only those tabular sources that contain explicit information about the presence or absence of sunspots
 then drawing sources appear to be consistent with the reliable tabular sources \citep{kovaltsov04,carrasco15}.

The main assumption in ZP15 is that sunspots were omitted, especially in verbal reports, if they were not round and
 did not resemble a planet.
The only direct example of that is given, with a reference to \citet{vaquero09}, that Harriot drew three sunspots on Dec~8 (=~Dec~18 greg.)
 1610 but wrote that the Sun was ``clear''.
However, this discussion was based on an incorrect interpretation of the original texts.
The actual statement of \citet{harriot1613} is
\begin{quote}
The altitude of the sonne being 7 or 8 degrees.
It being a frost and a mist.
I saw the sonne in this manner [drawing].
I saw it twise or thrise.
Once with the right ey and other time with the left.
In the space of a minute time, after the sonne was to cleare.
\end{quote}
As indicated by the observing times and numerous other statements about a ``well tempered'' Sun in
 the course of his observations, he mostly observed near sunrise or sunset or with a certain cloud
 cover to be able to look through the telescope.
The statement that ``the sonne was to cleare'' refers to the fact that the Sun became too
 bright after a few minutes of observing.
In this context, ``cleare'' means ``bright'' and not clear or spotless.
Therefore, this example was incorrectly taken by ZP15 as an illustrative case of the discrepancy between
 textual and drawing sources.

As another example, we compared the textual records by \citet{smogulecz_schoenberger1626}, who had conservative
 views on sunspots (see Sect.~\ref{Sec:ww}), with the drawings made by Scheiner in Rome for the same period of 1625.
We found that that Smogulecz and Sch\"onberger omitted a number of spots from the drawings, but mentioned
 all spots they saw in their text (calling them `stell\ae{}'), in accordance with Scheiner.
This is in contradiction with the assumption of ZP15 that verbal reports are subject to withholding spots.
Table~\ref{schoenberger_spots} lists the numbers of spots mentioned in their text versus those drawn in the figures.
(We note that the values are also incorrectly used in HS98.)
\citet{smogulecz_schoenberger1626} selected certain spots which were visible long enough to measure
 the obliquity of the Sun's axis with the ecliptic and plotted them schematically as circles as
 not particularly interested in their shape.
\begin{table}
\caption{Comparison of the number of spots listed in the verbal reports
versus the number of spots in the drawings by \cite{smogulecz_schoenberger1626}
in 1625.
\label{schoenberger_spots}}
\begin{tabular}{lrr|lrr}
\hline
Date        &  Text  & Drawn &Date        &  Text  & Drawn \\
\hline
1625 Jan 14 &   1    &  1 & 1625 Aug 22 &   2    &  1 \\
1625 Jan 15 &   1    &  1 & 1625 Aug 23 &   2    &  1 \\
1625 Jan 16 &   4    &  1 & 1625 Aug 27 &   6    &  1 \\
1625 Jan 17 &   8    &  1 & 1625 Aug 28 &  10    &  1 \\
1625 Jan 18 &   2    &  1 & 1625 Aug 31 &   7    &  1 \\
1625 Jan 19 &   4    &  1 & 1625 Sep 01 &   6    &  1 \\
1625 Jan 20 &   2    &  1 & 1625 Sep 05 &   8    &  4 \\
1625 Feb 12 &   8    &  1 & 1625 Sep 07 &   6    &  3 \\
1625 Feb 16 &  10    &  1 & 1625 Sep 08 &   6    &  3 \\
1625 Feb 17 &  11    &  1 & 1625 Sep 11 &   5    &  3  \\
1625 Feb 18 &  10    &  1 & 1625 Sep 12 &   4    &  3 \\
1625 Feb 21 &   4    &  1 & 1625 Sep 13 &   2    &  2 \\
1625 Jun 01 &   9    &  1 & 1625 Oct 05 &   9    &  8 \\
1625 Jun 04 &   3    &  1 & 1625 Oct 06 &   2    &  1 \\
1625 Jun 05 &   3    &  1 & 1625 Oct 09 &   4    &  4 \\
1625 Jun 06 &   2    &  1 & 1625 Oct 10 &   7    &  8 \\
1625 Jun 07 &   3    &  1 & 1625 Oct 11 &   9    &  9 \\
1625 Jun 09 &   2    &  1 & 1625 Oct 13 &   2    &  1 \\
1625 Aug 08 &   6    &  1 & 1625 Oct 14 &   2    &  1 \\
1625 Aug 09 &   4    &  1 & 1625 Oct 15 &   3    &  1 \\
1625 Aug 10 &   2    &  1 & 1625 Oct 25 &   1    &  1 \\
1625 Aug 12 &   4    &  2 & 1625 Oct 26 &   1    &  1 \\
1625 Aug 13 &   3    &  2 & 1625 Oct 27 &   1    &  1 \\
1625 Aug 14 &   3    &  2 & 1625 Oct 28 &   1    &  1 \\
1626 Aug 15 &   4    &  2 & 1625 Oct 29 &   1    &  1 \\
1625 Aug 17 &   2    &  2 & 1625 Oct 31 &   1    &  1 \\
1625 Aug 18 &   4    &  2 & 1625 Nov 01 &   1    &  1 \\
1625 Aug 19 &   2    &  1 &\\
\hline
\end{tabular}
\end{table}

\subsubsection{Relation between maximum number of sunspot groups and sunspot number}
\label{S:rel}

\begin{figure}[t]
\centering \resizebox{\columnwidth}{!}{\includegraphics{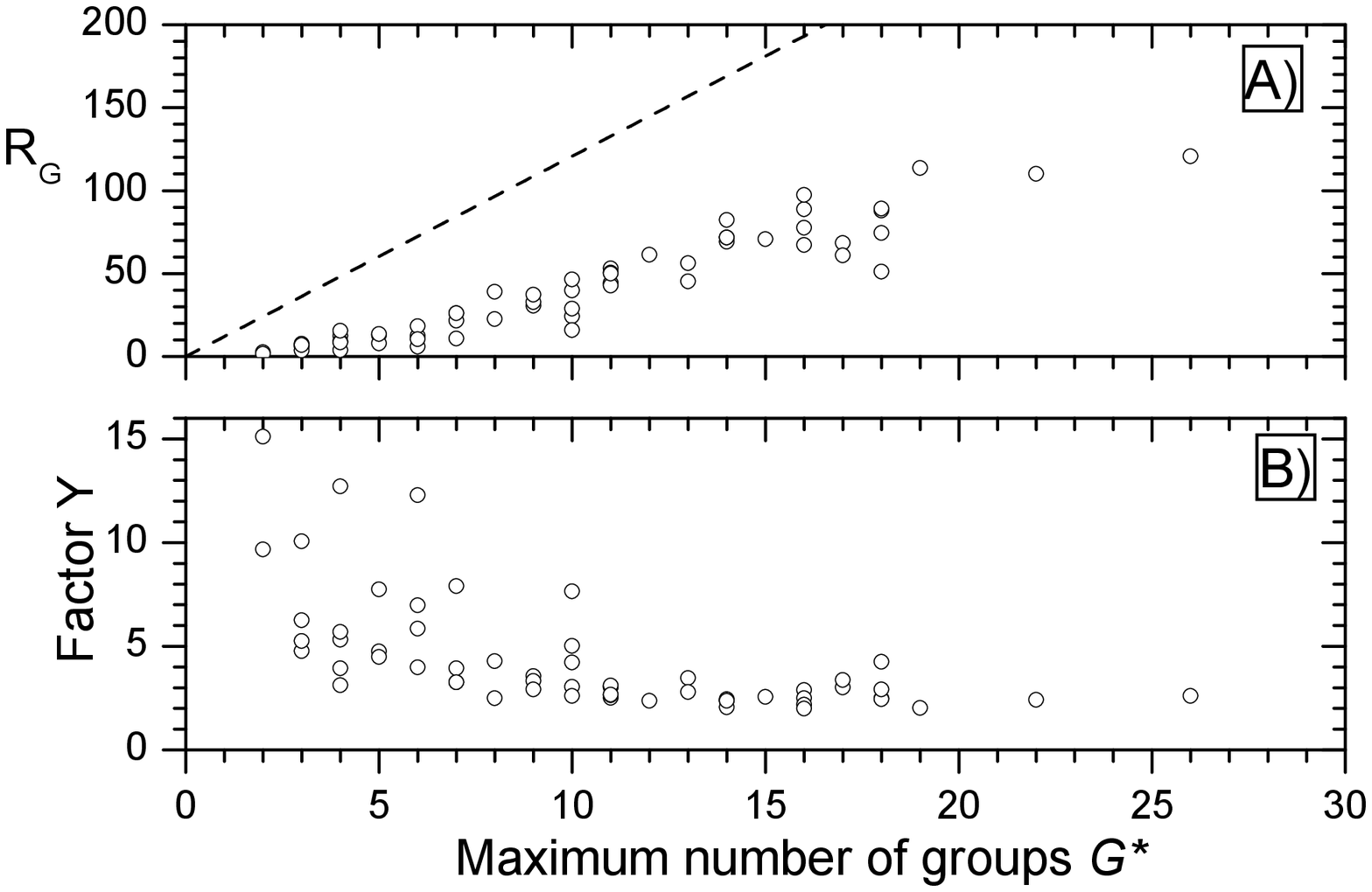}}
\caption{Illustration of the incorrectness of the method used by ZP15 to assess the group sunspot number $R_G$ during the MM.
Panel (A): Annual values of $R_G$ as a function of the maximum daily number of
sunspot groups $G^*$ for the same year in the HS98 catalogue
 for the period of 1886--1945; the dashed line is the dependence of $A^*_G = 12.08\times G^*$ used by ZP15.
Panel (B): The overestimate factor $Y$ of the GSN by the ZP15 method $Y=A^*_G/R_G$.
}
\label{Fig:G_Rg}
\end{figure}

ZP15 proposed a new method to assess the amplitude of the solar cycle during the MM.
As the amplitude of a sunspot cycle, $A_G^*$, they used the {\it maximum daily} number
 of sunspot groups $G^*$ during the cycle, so that $A^*_G = 12.08\times G^*$, where the coefficient
 12.08 is a scaling between the average number of sunspot groups and the sunspot number \citep{hoyt98}.
We note that using the maximum daily value of $G^*$ instead
 of the average value $G$ leads to a large overestimate of the sunspot cycle amplitude,
 particularly during the MM.
We analyze the HS98 database for the period 1886--1945, when sunspot cycles were not very high, to compare the
 annually averaged group sunspot numbers $R_G$ and the annual values of $A^*_G$ obtained using the annual maxima of the
 daily sunspot group numbers $G^*$.
Fig.~\ref{Fig:G_Rg}a shows a scatter plot of the annual values of $R_G$ and $G^*$ (dots), while the dashed line gives an estimate of the $A^*_G$
 based on $G^*$, following the recipe of ZP15.
One can see that while there is a relation between annual $R_G$ and $G^*$, the proposed method heavily
 overestimates the annual sunspot activity.
Fig.~\ref{Fig:G_Rg}b shows the overestimate factor $Y=A^*_G/R_G$ of the sunspot numbers as a function of $G^*$.
While the factor $Y$ is 2--3 for very active years with $G^*>15$, the overestimate can reach an order
 of magnitude for years with weaker activity, such as during the MM.
When applying to the sunspot cycle amplitude, the error becomes even more severe.
Thus, by taking the cycle maxima of the daily number of sunspot groups instead of their annual means,
 ZP15 systematically overestimated the sunspot numbers during the MM by a factor of 5--15.

{\it The number of sunspot groups in 1642.}

\noindent
ZP15 proposed that the solar cycle just before the MM was high (sunspot number $\approx 100$) which is based on
 a report of 8 sunspot groups observed by Antonius Maria Schyrleus of Rheita in February 1642 as presented in the HS98 database.
However, as shown by \citet{gomez15}, this record is erroneous in the HS98 database because it is based on an incorrect
 translation from the original Latin records, which says that one (or a few) group was observed for 8 days in
 June 1642 instead of 8 groups in February 1642.
Accordingly, the maximum daily number of sunspot groups reported for that cycle $G^*$ was 5, not 8, reducing the
 cycle amplitude claimed by ZP15 (see Sect.~\ref{S:rel}) by about 40\%.

{\it The number of sunspot groups in 1652.}

\noindent
The original HS98 record contains 5 sunspot groups for the day of Apr-01-1652, referring to observations by Johannes Hevelius.
Accordingly, this value (the highest daily $G^*$ for the decade) was adopted by ZP15 leading to the
 high proposed sunspot cycle during the 1650s.
However, as discussed by \citet{vaquero_SP_14} in great detail, this value of 5 sunspot groups is an erroneous interpretation, by HS98 with
 reference to \citet{wolf1856}, of the original Latin text by Hevelius saying that there were 5 spots in two distinct groups on the Sun.
Accordingly, the correct value of $G^*$ for that day should be 2 not 5.

{\it The number of sunspot groups in 1705.}

\noindent
A high sunspot number of above 70 was proposed by ZP15 for the year 1705 based on six sunspot groups reported by J. Plantade from Montpellier
 (the correction factor for this observer is 1.107 according to HS98) for the day Feb-13-1705.
This observer was quite active with regular observations during that period, with 44 known daily observational reports for
 the year 1705.
For example, J. Plantade reported 2, 3, 6, and 1 groups, respectively, for the days of Feb-11 through Feb-14.
His reports also mention the explicit absence of spots from the Sun after the group he had followed passed beyond the limb.
However, he did not make any reports during long spotless periods, and wrote notes again when a new sunspot group appeared.
The average number of sunspot groups per day reported by J. Plantade for 1705 was 1.22, which is a factor of 5 lower than that adopted
 by ZP15 who only took the largest daily value (see Sect.~\ref{S:rel}).
If one calculates the group sunspot number from the dataset of J. Plantade records for 1705 in the classical way,
 one obtains a value of $R_G=16.3$ ($1.22\times 12.08\times 1.107$).

\subsection{Butterfly diagram}
\label{S:lat}
\begin{figure}[t]
\centering \resizebox{\columnwidth}{!}{\includegraphics{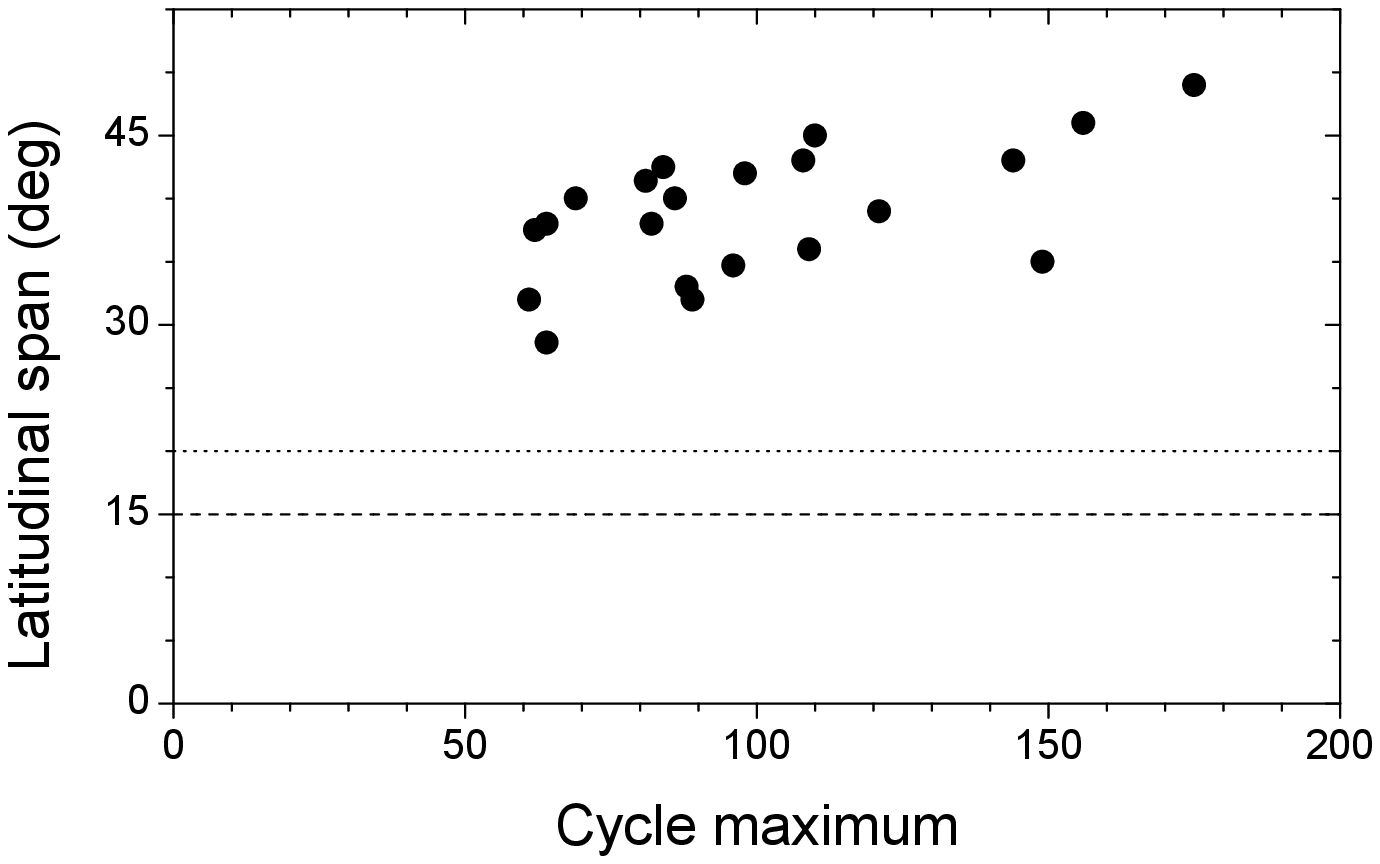}}
\caption{Maximum latitudinal span of the butterfly diagram as a function of the cycle amplitude in annual $R_G$ for
 solar cycles \# 0--4 and 7--22, accounting only for large spots (with area greater than 100 msd).
The dashed and dotted lines depict the maximum latitudinal extent of sunspot occurrence during the core (1645--1700) and the
 entire MM (1645--1712).
}
\label{Fig:SN_lat}
\end{figure}

According to sunspot drawings during some periods of the MM, a hint of the butterfly diagram has been identified,
 particularly towards the end of the MM after 1670 \citep{ribes93,soon03,casas06}.
However, the latitudinal extent of the butterfly wings was quite narrow, being within $15^\circ$ for the core MM
 (1645--1700) and $20^\circ$ for the period around 1705, while cycles before and after the MM had a latitudinal extent
 of $30^\circ$ or greater.
This suggests that the sunspot occurrence during the MM was limited to a more narrow band than outside the MM.
Here we compare the statistics of the latitudinal extend of the butterfly diagram wings for solar cycles \# 0 through 22
 (cycles 5 and 6 are missing).
The cycles \#0 through 4 were covered by digitized drawings made by Staudacher for the period 1749--1792 \citep{arlt08},
  cycles \#7 through 10 (1825--1867) were covered by digitized drawings made by Samuel Heinrich Schwabe \citep{arlt13},
  cycle \# 8 by drawings of Gustav Sp\"orer \citep{diercke15}, while the period after 1874 was studied using the Royal
  Greenwich Observatory (RGO) catalogue.
Moreover, a machine-readable version of the sunspot catalogues of the 19th century complied by Carrington, Peters
 and de la Rue has been released recently \citep{casas14}.
For each solar cycle we defined the maximum latitude (in absolute values without differentiating north and south) of sunspot
 occurrence.
Since the telescopic instruments were poorer during the MM than nowadays, for consistency we considered only large spots
 with the projected spot area greater than 100 msd (millionths of the solar disc).
The result is shown in Fig.~\ref{Fig:SN_lat} as a function of the cycle maximum (in $R_G$).
One can see that there is a weak dependence for stronger cycles generally having a larger latitudinal
 span \citep[cf, e.g.][]{vitinsky86,solanki08,jiang11},
 but the latitudinal extent of the butterfly wing was always greater than 28$^\circ$ for the last 250 years.
A robust link between the mean/range latitude of sunspot occurrence and cycle strength is related to the dynamo
 wave in the solar convection zone and has been empirically studied, e.g., by \citet{solanki08} or \citet{jiang11_2}.
Since the maximum latitudinal extent of sunspots during the MM was 15$^\circ$ (during the core MM) or 20$^\circ$ (around ca. 1705),
 it suggests a weak toroidal field causing a narrower latitudinal range of sunspot formation during the MM.
This conclusion is in agreement with the results of a more sophisticated analysis by \citet{ivanov11} who found that
 the latitudinal span of the butterfly diagram during the late part of the MM should be 15--20$^\circ$, i.e.,
 significantly lower than during the normal cycles.
One may assume that all the higher latitude spots were deliberately omitted by all the observers during the MM but we
 are not aware of such a bias.

We note that two data sets of sunspot latitudes during the MM have been recently recovered and translated
 into machine-readable format \citep{vaquero_ASR_15}.
Using these data sets, a decadal hemispheric asymmetry index has been calculated confirming a very strong hemispherical
 asymmetry (sunspots appeared mostly in the southern hemisphere) in the MM, as reported in earlier works \citep{spoerer1889,ribes93,sokoloff94}.
Another moderately asymmetric pattern was observed only in the beginning of the Dalton minimum \citep{arlt09,usoskin_arlt_09}.
Thus, this indicates that the MM was also a special period with respect to the distribution of sunspot latitudes.

\subsection{East-Asian naked-eye sunspot observations}
\label{S:naked}
East-Asian chronicles reporting observations for about two millennia, by unaided naked eyes, of phenomena
that may be interpreted as sunspots have sometimes been
 used as an argument suggesting high solar activity during the MM
 \citep{schove83,nagovitsyn01,ogurtsov03,zolotova15}.
Such statements are based on an assumption that sunspots must be large to be observed and that
 this is possible only at a high level of solar activity.
However,  as shown below, this is not correct.
While such historical records can be useful in a long-term perspective showing
 qualitatively the presence of several Grand minima during
 the last two millennia \citep{clark78,vaquero02,vaquero09} including also the MM,
 this dataset is not useful for establishing the quantitative level
 of solar activity over short timespans due to
 the small number of individual observations and/or the specific meteorological, sociological
 and historical conditions required for such record \citep[see Chapter 2 in][]{vaquero09}.
Moreover, it is very important to indicate that the quality of the historical
 record of naked-eye sunspot (NES) observations
 was not uniform through the ages (i.e. during the approximately two milliennia covered by the record).
In fact, the quality of such records for the last four centuries was much poorer
 than that for the 12th-15th centuries, due to a change in the type of historical sources.
In particular, the data coverage was reduced greatly after 1600 \citep[see Figure 2.18 in][]{vaquero09}.
There are very few NES records during the century between the MM and the Dalton
 Minimum, representing the social conditions supporting such observations and the maintenance of such records
 rather than sunspot activity.
Therefore, the historical record of NES observations is not useful to estimate the level
 of solar activity during recent centuries \citep{eddy83,mossman89,willis96}.

\subsubsection{Do NES observations imply high activity?}

It is typical to believe that historical records of NES observations necessarily imply very high levels of solar activity
 \citep[e.g.,][]{ogurtsov03}, assuming that observable spots must have a large area exceeding 1900 msd (millionths of the solar disc)
 with a reference to \citet{wittmann78}.
However, the latter work does not provide any argumentation for such a value and, as shown below, this is not correct.

Here we show that reports of NES observations do not necessarily correspond to high activity or even to big spots.
We compared the East-Asian sunspot catalogue by \citet{yau_steph88} for the period 1848--1918 (25 reported
 naked-eye observations during 21 years) with data from the HS98 catalogue.
\begin{figure}[t]
\centering \resizebox{\columnwidth}{!}{\includegraphics{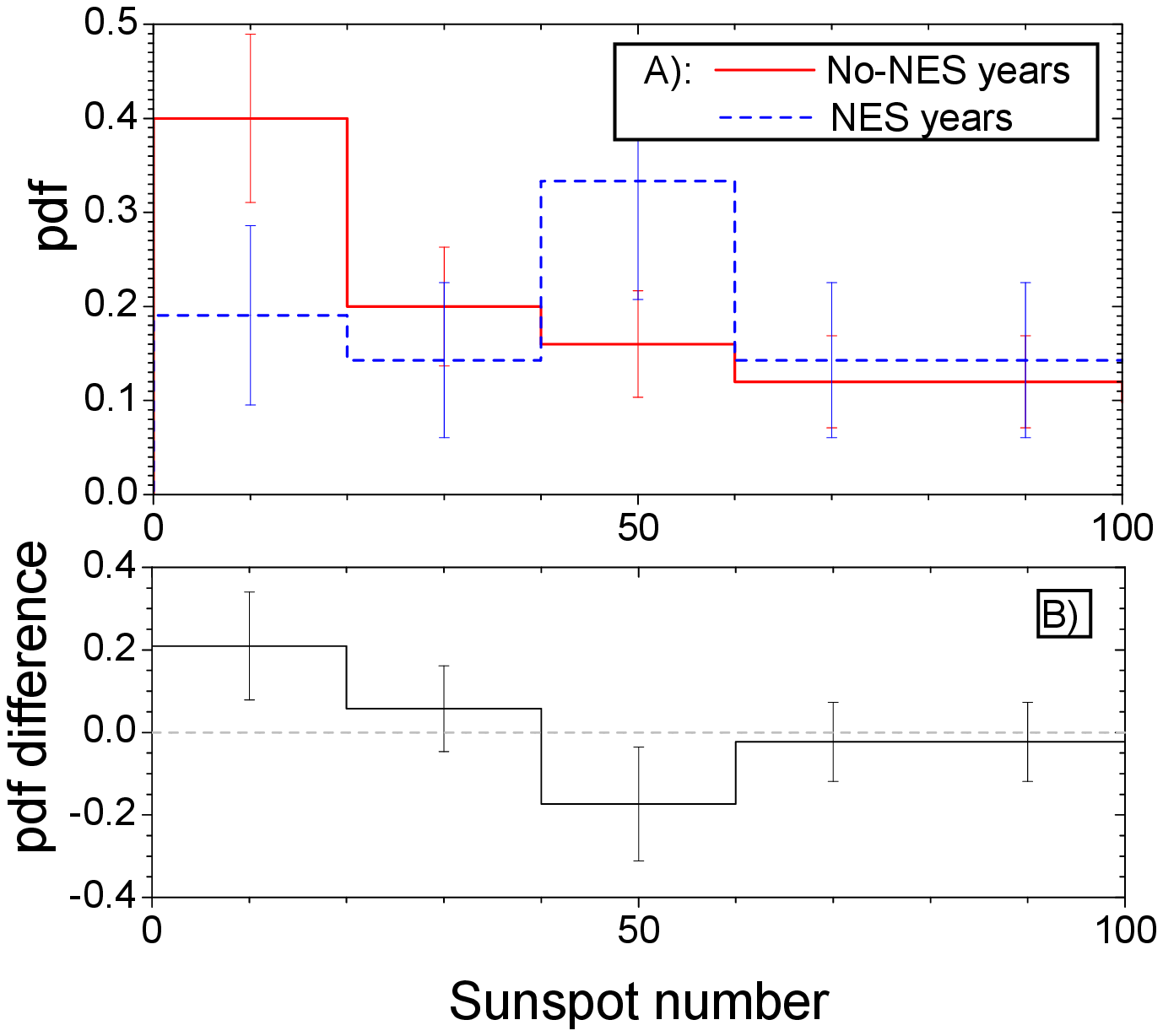}}
\caption{Probability density function for occurrence of the annual group sunspot numbers for the years 1848--1918.
Panel A: The red solid line represents the years (50 years) without naked-eye spot (no-NES) reports, while the blue dotted line represents
 only the years (21 years) with NES.
Panel B: The difference between the no-NES and NES probability density functions.
Error bars represent the $1\sigma$ statistical uncertainties.
}
\label{Fig:naked}
\end{figure}
Figure~\ref{Fig:naked} shows the probability density functions (pdf's) of the
 sunspot numbers for the years with and without NES observations.
One can see that the probability of NES reports to occur does not depend on the actual sunspot number
 as the blue dotted curve in panel A is almost flat, while intuitively it should be expected
 to yield larger probability for high sunspot numbers and to vanish for small sunspot numbers.
Moreover, there is no statistically significant difference in the sunspot numbers between
 the two pdfs.
Accordingly, the null hypothesis that the two pdfs belong to the same population
 cannot be statistically rejected.
Obviously, there is no preference to NES
 observations during the years of high sunspot numbers.
The naked-eye reports appear to be distributed randomly, without any
 relation to the actual sunspot activity.
Accordingly, the years with unaided naked-eye sunspot reports provide no
 preference for the higher sunspot number.

\begin{figure}[t]
\centering \resizebox{\columnwidth}{!}{\includegraphics{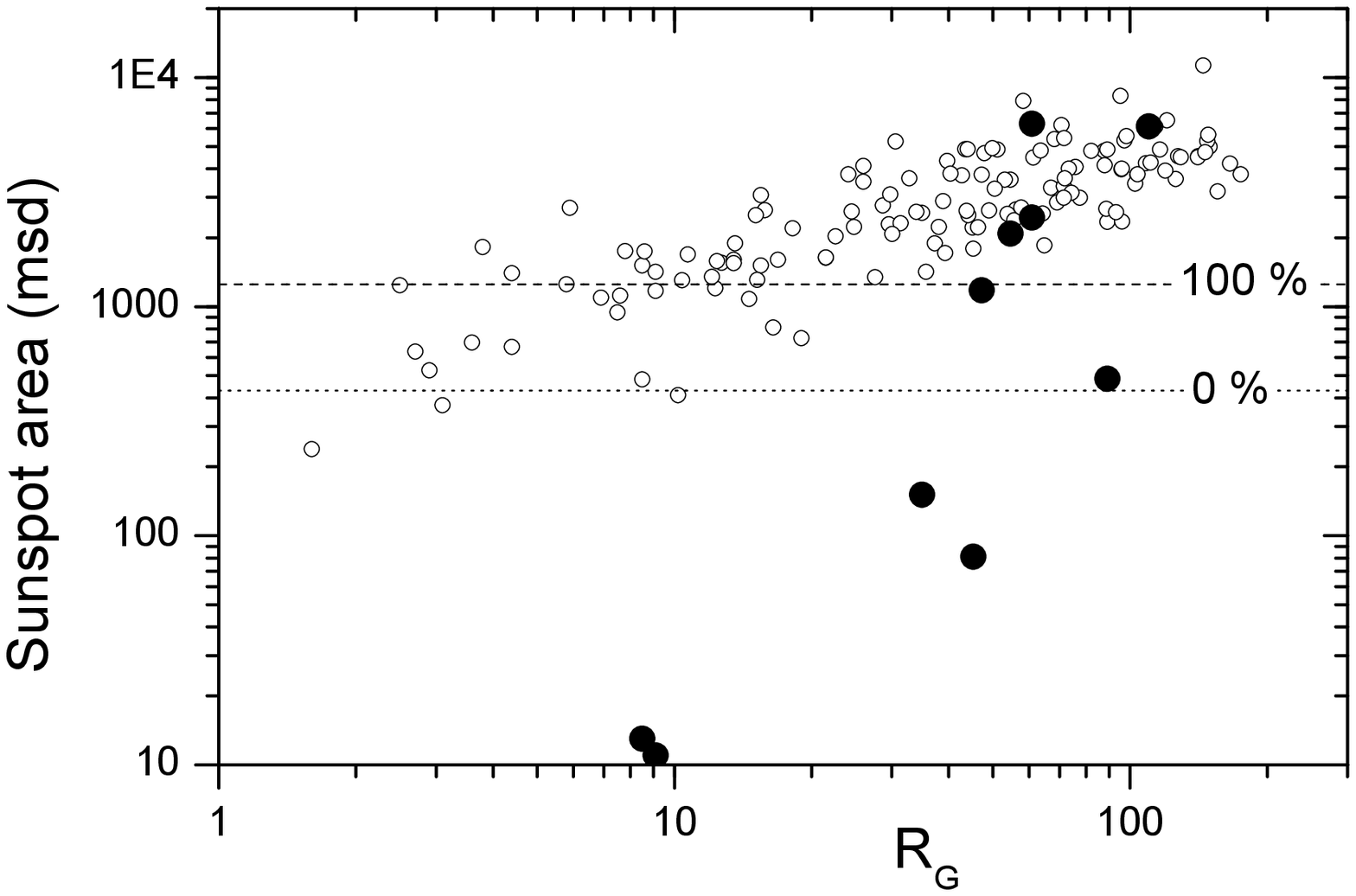}}
\caption{Open dots depict dependence of the area of the largest sunspot within a year (GSO data)
 vs. the annual group (1874--1996) or international (1997--2013) sunspot numbers.
 Big filled dots denote the largest sunspot's area for the days of reported naked-eye sunspot observations during
 the period 1874--1918 \citep{yau_steph88}.
 The dashed and dotted lines depict the 100\% (all spots above this line are visible) and 0\% (detectability threshold)
 probability of observing a sunspot of the given area by an unaided eye, according to \citet{schaefer93} and \citet{vaquero09}.}
\label{Fig:area}
\end{figure}

Next we study the correspondence between the NES reports and actual sunspots during the exact dates
 of NES observations (allowing for 1 day dating mismatch because of the local time conversion).
The data on the sunspot area were taken from the Royal Greenwich Observatory (RGO)
sunspot group photographic catalogue
 (http://solarscience.msfc.nasa.gov/greenwch.shtml).
Figure~\ref{Fig:area} shows, as filled circles, the largest observed sunspot area for the days when East-Asian NES
 observation were reported during the years 1874--1918 \citep{yau_steph88}.
Detectability limits of the NES observations \citep{schaefer93,vaquero09} are
shown as dotted (no spots smaller than $\approx 425$ msd can be observed by
 the unaided eye) and dashed (all spots greater
  than $\approx 1240$ msd are observable) lines.
One can see that half of the reported NESs lie below or at the lower detectability limit and are
 not visible by a normal unaided human eye, likely being spurious or
 misidentified records \citep[cf.][]{willis96}.

As an example we consider two dates with NES records with the smallest sunspots.
A sunspot was reported to be seen by naked eyes on Feb-15-1900, when there
 were no sunspots on the Sun according to RGO, while
 there was one very small group (11 msd area) on the pervious day of Feb-14-1900.
Another example of a NES report is for the day of Jan-30-1911 when there was
a single small group (area 13 msd) on the Sun
 \citep[see also Fig.~9 in][]{yau_steph88}.
Such small groups cannot be observed by an unaided eye.
Moreover, in agreement with the above discussion, even for big spots above the 100\% detectability level,
 the relation to solar activity is unclear.
Open dots in Figure~\ref{Fig:area} denote the area of the largest spot observed each year
 vs. the mean annual sunspot number for years 1874--2013.
One can see that the occurrence of a large sunspot detectable by naked eye
 does not necessarily correspond
 to a high annual sunspot number, as it can occur at any level of solar activity from $R_G=3$ to 200.

\begin{figure}[t]
\centering \resizebox{\columnwidth}{!}{\includegraphics{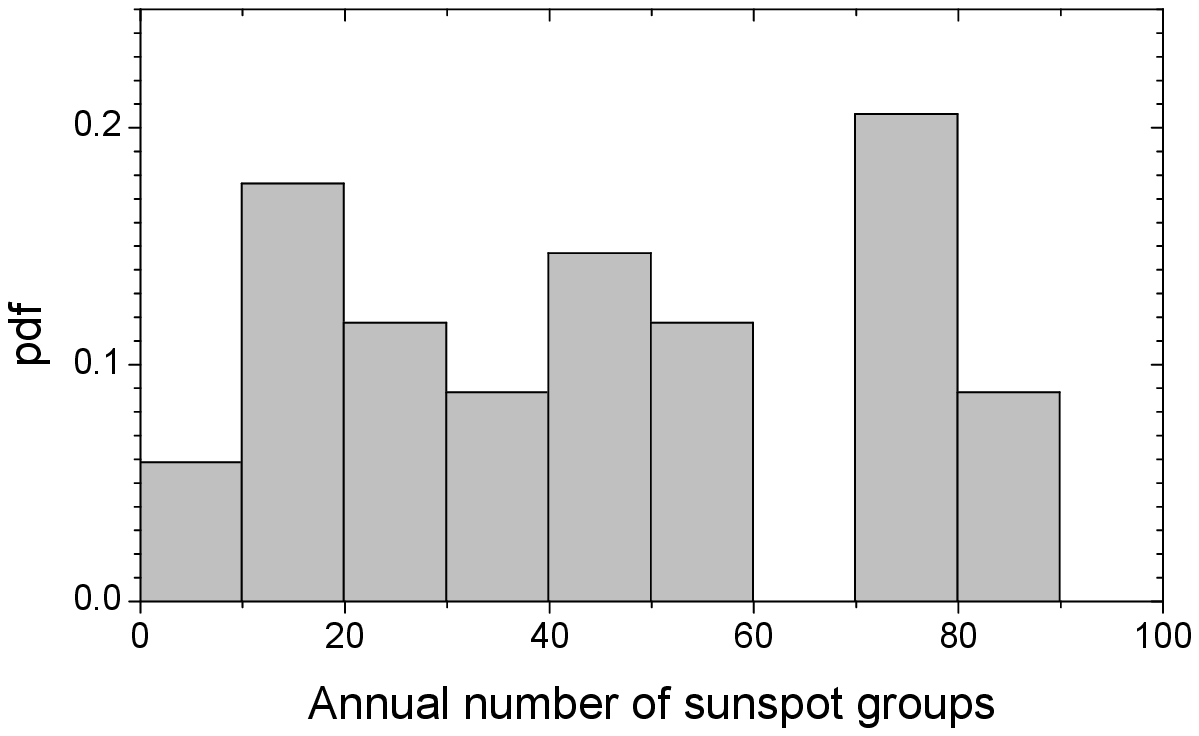}}
\caption{Probability density function of the  occurrence of NES records by S.H. Schwabe as a function of the annual groups sunspot number
 during the years of Schwabe's observations.}
\label{Fig:sch}
\end{figure}

Another data set is provided by the naked-eye observations by Samuel Heinrich Schwabe, who recorded telescopic sunspot
 data in 1825--1867, but also occasionally reported on naked-eye visibilities of sunspot groups (Pavai et al., 2015, in prep.).
We analyzed the (annual) group sunspot numbers for each event when Schwabe reported a naked-eye visibility, as
 shown in Fig.~\ref{Fig:sch} in the form of a probability density function versus the annual group sunspot number.
The NES reports were quite frequent during years with low sunspot activity ($\approx 25$\% of naked eye spots were
 reported for the years with $R_G$ below 20, some of them even below 10).
It is interesting that about 20\% of naked eye observations by Schwabe were reported on days with a single group
 on the solar disc.
We note that Schwabe certainly looked without the telescope when he saw a big group with it, so the selection may
 be biased towards larger spots.
On the other hand, it is unlikely that he would watch out for naked-eye spots only if there were just few group on the Sun,
 therefore we do not expect an observational bias towards lower activity periods.

Thus, a significant part of the East-Asian NES observational reports are unlikely to be real observations and, even if they
 were correct, they do not imply a high level of solar activity.
This implies that the NES reports cannot be used as an index of sunspot activity in a simple way
 \citep[cf.,][]{eddy83,willis96,mossman89,usoskin_LR_13}.

\subsubsection{NES observations around the MM}

According to the well-established catalogue of NESs \citep{yau_steph88} from Oriental chronicles, NESs were observed relatively
 frequently before the MM -- 16 years during the period 1611--1645 are marked with the NES records.
A direct comparison between the NES catalogue and the HS98 database \citep[with the correction by][]{vaquero11}
 shows that the NES records either are confirmed by Eurpean telescopic observations (Malapert, Schenier, Mogling, Gassendi, Hevelius)
 or fall in data gaps (after removing generic statements from the HS98 database).
There is no direct contradictions between the datasets for that period.

There are several NES records also during the MM but they are more rare (8 years during 1645--1715), as discussed in detail here.

Three NES observations are reported for the years 1647, 1648 and 1650, respectively, which fall in a long gap (1646--1651)
 of telescopic observations where only a generic statement by Hevelius exists.
The exact level of sunspot activity during these years is therefore unknown.

A NES report dated Apr 30 (greg.) 1655 falls in a small gap in the HS98 database but there is some activity reported
 in the previous month of March.
The mean annual GSN for the year 1655, estimated in the `loose' model of \citet{vaquero15}, is $R_{\rm G}=5.7$.

A NES was reported in the Spring of 1656 which overlaps with a sunspot group reported by Bose in February.
$R_{\rm G}$ is estimated in the `strict' model as 12.7 \citep{vaquero15}.

There are four NES records for the year 1665, but three of them are likely to be related to the same event in late
 February, and one to Aug 27, thus yielding two different observations.
These events again fall into gaps in direct telescopic observations with only generic statements available.
For this year, only nine daily direct telescopic records, evenly spread over the second half-year, exist.
The observers Hevelius and Mezzavacca, both claimed the absence of spots.
The exact level of activity for this year is therefore unknown.
Probably, there was some activity in 1665 but not high, owing to the direct no-spots records (cf. Sect.~\ref{S:1660}).

Another NES was reported to be observed for three days in mid-March 1684, which falls on no-spot records by la Hire.
We note that this year was well covered by telescopic observation, especially in the middle and late year, and it was
 relatively active $R_{\rm G}=11.7$ \citep[][ `strict' model]{vaquero15}.
Account for the probably missed spot in March would raise the annual GSN value of this otherwise well observed year
 by less than 2.

One more NES record is for the year 1709 (no date or even season given).
That year was well observed by different observers, with some weak activity reported intermittently throughout the year.
The mean annual GSN in the `strict' model is $R_{\rm G}=5.3$.

Thus, except for the year 1684, there is no direct clash between the East-Asian NES records and European
 telescopic observations, and the former do not undermine the low level of solar activity suggested by the latter.

%_________________________________
\section{Indirect proxy data}
\label{S:proxy}

\subsection{Aurorae borealis}
\label{S:aurorae}
\subsubsection{Geomagnetic Observations}
In recent years we have learned a great deal from geomagnetic observations about centennial-scale solar variability
 and how it influences the inner heliosphere, and hence the Earth \citep{lockwood99,lockwoodLR}.
Such studies cannot tell us directly about the MM because geomagnetic activity was first observed in
 1722 by George Graham in London and the first properly-calibrated magnetometer was not introduced until 1832 (by Gauss in G\"ottingen).
Graham noted both regular diurnal variations and irregular changes during the peak of solar cycle \# -3 (ca. 1720), which was
 the first significant cycle after the MM.
This raises an interesting question: were these observations made possible by Graham's advances to the compass
 needle bearing and observation technique or had magnetic activity not been seen before because it had not been strong enough?
However, despite coming too late to have direct bearing on understanding the MM, the historic geomagnetic
 data have been extremely important because they have allowed us to understand and confirm the link between sunspot
 numbers and cosmogenic isotope data.
In particular, they have allowed modelling of the open solar flux which shows that the low sunspot numbers in
 the MM are quantitatively (and not just qualitatively) consistent with the high cosmogenic isotope abundances
 \citep{solanki00,owens12,lockwood_3_14}.
This understanding has allowed the analysis presented in section~\ref{S:cosmo}.

\subsubsection{Surveys of historic aurorae}
Earlier in the same solar cycle as Graham's first geomagnetic activity observations, on the night of Tuesday 17th March 1716
 (Gregorian calendar: note the original paper gives the Julian date in use of the time which was 6th March),
 auroral displays were seen across much of northern Europe, famously reported by Edmund \citet{halley1716} in Great Britain.

What is significant about this event is that very few people in the country had seen an aurora before \citep{fara96}.
Indeed, Halley's paper was commissioned by the Royal Society for this very reason.
This event was so rare it provoked a similar review under the auspices of  l'Acad\'emie des Sciences of Paris (by Giacomo
 Filippo Maraldi, also known as Jacques Philippe  Maraldi) and generated interest at the Royal Prussian Academy of Sciences in Berlin (by Gottfried
 Wilhelm Leibnitz).
All these reviews found evidence of prior aurorae, but none in the previous half century.

Halley himself had observed the 1716 event (and correctly noted that the auroral forms were aligned by the magnetic field) but
 had never before witnessed the phenomenon.
It is worth examining his actual words: ``...[of] all the several sorts of meteors [atmospheric phenomena] I have hitherto
 heard or read of,  this [aurora] was the only one I had not as yet seen, and of which I began to despair, since it is certain
 it hath not happen'd to any remarkable degree in this part of England since I was born [1656]; nor is the like recorded in the
 English Annals since the Year of our Lord 1574.''
This is significant because Halley was an observer of astronomical and atmospheric phenomena
 who even had an observatory constructed in the roof of his house
 in New College Lane, Oxford where he lived from 1703 onwards.
In his paper to the Royal Society, Halley lists reports of the phenomenon, both from the UK and abroad, in the years 1560,
 1564, 1575, 1580, 1581 (many of which were reported by Brahe in Denmark), 1607 (reported in detail by Kepler in Prague)
 and 1621 (reported by Galileo in Venice and Gassendi in Aix, France).
Strikingly, thereafter Halley found no credible reports until 1707 (R{\o}mer in Copenhagen and Maria and Gottfried Kirch in Berlin) and 1708
 (Neve in Ireland).
He states ``And since then [1621] for above 80 years, we have no account of any such sight either from home or abroad''.
This analysis did omit some isolated sightings in 1661 from London (reported in the Leipzig University theses by Starck and Fr\"uauff).
In addition to being the major finding of the reviews by Halley, Miraldi and others (in England, France and Germany), a similar
 re-appearance of aurorae was reported in 1716-1720 in Italy and in New England \citep{siscoe80}.

\begin{figure}[t]
\centering \resizebox{\columnwidth}{!}{\includegraphics{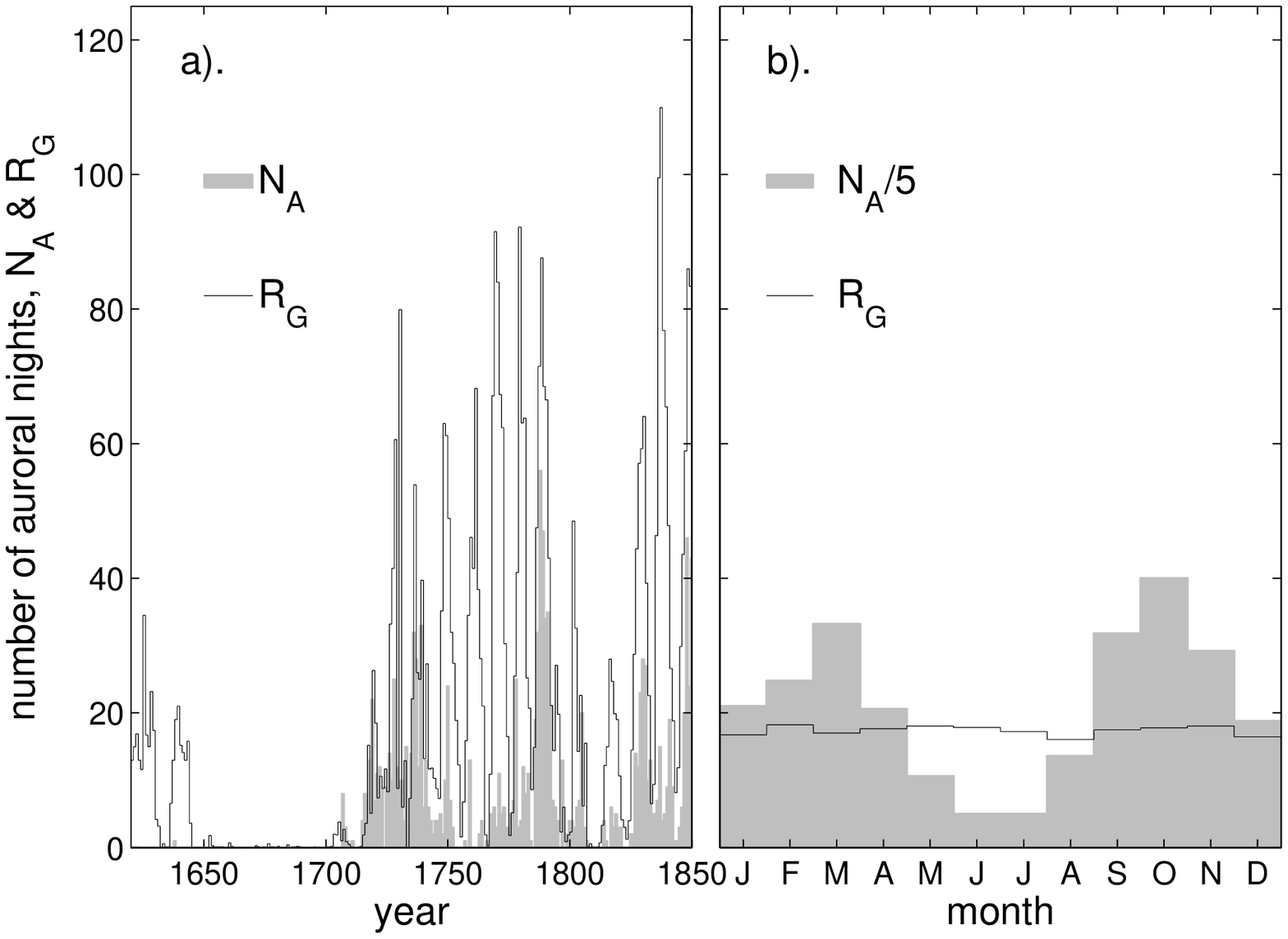}}
\caption{(a) The grey histogram shows the number of auroral nights, $N_A$, in calendar years for observations in Great Britain
 collated by E.J. \citet{lowe1870} with the addition of the observations by Thomas Hughes \citep{harrison05} and John Dalton \citep{dalton1834}.
The black line shows the annual group sunspot number of \citet{hoyt98}, with the
 adoption of recent corrections by \citet{vaquero11} and \citet{vaquero_SP_14}.
 Lowe's personal copy of his catalogue of natural phenomena (including auroras) was only recently discovered and was compiled completely
  independently of other catalogues.
 Yet it shows, like the others, the dearth of sightings during the Maunder minimum, some events in 1707 and 1708 and the return
 of regular sightings in 1716.
 (b) Annual variation of $N_A$ in the same dataset and of $R_G$.
}
\label{Fig:lowe}
\end{figure}

The absence of auroral sightings in Great Britain during the MM is even more extraordinary when one
 considers the effects of the secular change in the geomagnetic field.
For example, using a spline of the IGRF (International Geomagnetic Reference Field, http://www.ngdc.noaa.gov/IAGA/ vmod/igrf.html)
 model after 1900 with the gufm1 model \citep{jackson00} before 1900 we find the geomagnetic latitude of Halley's observatory
 in Oxford was 60.7$^\circ$ in 1703 and Edinburgh was at 63.4$^\circ$.
Auroral occurrence statistics were taken in Great Britain between 1952 and 1975, and of these years the lowest annual mean sunspot number
 was 4.4 in 1954.
Even during this low solar activity year there were 169 auroral nights observed at the magnetic latitude that Edinburgh had during
 the MM and 139 at the magnetic latitude that Oxford had during the MM \citep{paton58}.
In other words, The British Isles were at the ideal latitudes for observing aurora during the MM and yet the number reported
 was zero.
This is despite some careful and methodical observations revealed by the notebooks of several scientists: for example, Halley's
 notebooks regularly and repeatedly use the term ``clear skies'' which make it inconceivable that he would not have noted an aurora had it been present.

Halley's failure to find auroral sightings in the decades before 1716 is far from unique.
Figure~\ref{Fig:lowe} is a plot of auroral occurrence in Great Britain from a previously unknown source.
It is shown with the group sunspot number $R_G$ during the MM.
This catalogue of auroral sightings in the UK was published in 1870 by an astronomer and a Fellow of the Royal Society, E.J. Lowe,
 who used parish records, newspaper reports and observations by several regular observers.
His personal copy of the book (with some valuable "marginalia" - additional notes written in the margin) was
 recently discovered in the archives of the Museum of English Rural Life at the University of Reading, UK \citep{lowe1870}.
Here we refer to this personally commented copy of the book.
We here have added to Lowe's catalogue of English recordings the observations listed in the diary of Thomas Hughes
in Stroud, England \citep{harrison05} and the observations made by John Dalton in Kendall and later Manchester \citep{dalton1834}.
Like so many other such records, the time series of the number of auroral nights during each year, shown by the grey
 histogram in Figure~\ref{Fig:lowe}a, reveals a complete dearth of auroral sightings during the MM.
As such, this record tells us little that is not known from other surveys; however, it is important to note that this
 compilation was made almost completely independently of, and using sources different from other catalogues such as those by
 de Mairan and Fritz (see below).

Figure~\ref{Fig:lowe}b shows the annual variation of the number of auroral nights and reveals the
 semi-annual variation \citep{Sabine_1852} \citep[equinoctial peaks in auroral occurrence were noted by][]{demairan1733}.
A corresponding semi-annual variation in geomagnetic activity \citep{Sabine_1852,cortie1912} is mainly caused by the effect of solar illumination of
 the nightside auroral current electrojets \citep{Cliver_2000, Lyatsky_etal_2001, Newell_etal_2002}, leading to equinoctial maxima
 in geomagnetic activity.
On the other hand, lower-latitude aurorae are caused by the inner edge of the cross-tail current sheet being closer to
 the Earth, caused by larger open flux in the magnetosphere-ionosphere system \citep{lockwoodLR} and so are more likely to
 be caused by the effect of Earth's dipole tilt on solar wind-magnetosphere coupling and, in particular the magnetic
 reconnection in the magnetopause that generates the open flux \citep{russel73}.
This is convolved with a summer-winter asymmetry caused by the length of the annual variation in the dark interval
 in which sightings are possible.
Note that Figure~\ref{Fig:lowe}b shows a complete lack of any annual variation in group sunspot number, as expected.
This provides a good test of the objective nature of the combined dataset used in Figure~\ref{Fig:lowe}.
Both parts (a) and (b) of Figure~\ref{Fig:lowe} are very similar in form to the corresponding plots made using
 all the other catalogues.

Elsewhere, however, other observers in 1716 were familiar with the phenomenon of aurorae \citep{brekke94}.
For example Joachim Ramus, a Norwegian (born in Trondheim in 1685 but by then living in Copenhagen), also witnessed aurora
 in March 1716, but unlike Halley was already familiar with the phenomenon.
Suno Arnelius in Uppsala had written a scientific thesis on the phenomenon in 1708.
Indeed after the 1707 event R{\o}mer had noted that, although very rare in his native Copenhagen, such events were
 usually seen every year in Iceland and northern Norway (although it is not known on what basis he stated this) \citep{stauning11,brekke94}.
But even at Nordic latitudes aurorae appear to have been relatively rare in the second half of the 17th century \citep{brekke94}.
Petter Dass, a cleric in Alstahaug, in middle Norway, who accurately and diligently reported all that he observed
 in the night sky between 1645 and his death in 1707, and had read many historic reports of aurorae, never once
 records seeing it himself.
In his thesis on aurorae (completed in 1738), Peter M{\o}øller of Trondheim argues that the aurora reported over
 Bergen on New Year's Eve 1702 was the first that was ever recorded in the city.
Celsius in Uppsala was 15 years old at the time of the March 1716 event but subsequently interviewed many older
 residents of central Sweden and none had ever seen an aurora before.
Johann Anderson was the mayor of Hamburg and discussed aurorae with Icelandic sea captains. They told him that
 aurorae were seen before 1716, but much less frequently \citep[reported in][]{horrebow1752}.
An important contribution to the collation of reliable auroral observations was written in 1731 by Jean-Jacques d'Ortous
 de Mairan \citep{demairan1733}, with a second edition published in 1754.
Both editions are very clear in that aurorae were rare for at least 70 years before their return in 1716.
The more thorough surveys by \citet{lovering1860}, \citet{fritz1873,fritz1881} and \citet{link64,link78} have all
 confirmed this conclusion \citep[see][]{eddy76}.

%All catalogues of historic auroral sightings are plagued by inaccurate reporting of dates (leading to double
% reporting of "an auroral night") and embellishment by historians aiming at adding drama to political, societal or military events.
%It has even been argued that some auroral reports were invented to generate public fear and so aid the collection of taxes and tithes.
%In addition, clouds add a chance element to their detection and sunlight makes observing seasons and times of day shorter
% at higher latitudes: a particular problem for the European sector where a given magnetic latitude is at a higher geographic latitude
% than at other longitudes.
%As surveys have been repeated, the anecdotal and debatable cases have gradually been removed and recent surveys have been able
% to use verifiable reports \citep{vazquez14}.

\subsubsection{Reports of aurorae during the Maunder minimum}

The above does not mean that auroral sighting completely ceased during the Maunder minimum.
de Mairan's original survey reported 60 occurrences of aurorae in the interval 1645-1698.
Many authors have noted that the solar cycle in auroral occurrence continued during the Maunder minimum
 \citep{link78,vitinsky78,gleissberg77,schroder92, legrand92}.
One important factor that must be considered in this context is the magnetic latitude of the observations.
It is entirely possible that aurora were always present, at some latitude and brightness, and that the main variable with
 increasing solar activity is the frequency of the equatorward excursions of brighter forms of aurorae.
In very quiet times, the aurora would then form a thin, possibly fainter, band at very high latitudes, with greatly
 reduced  chance of observation.
An important indication that this was indeed the case comes from a rare voyage into the Arctic during the MM
 by the ships Speedwell and Prosperous in the summer of 1676.
This was an expedition approved by the then secretary to the British Admiralty, Samuel Pepys, to explore the
 north east passage to Japan.
Captained by John Wood, the  ships visited northern Norway and Novaya Zemlya (an Arctic archipelago north of Russia),
 reaching a latitude of 75$^\circ$ 59' (geographic) before the Speedwell ran aground.
Captain Wood reported that aurorae were only seen at the highest latitudes by the local seaman that he met.
Furthermore, anecdotal evidence was supplied by Fritz who quoted a book on Greenland fisheries that aurorae were sometimes
 seen in the high Arctic at this time.
The possibility of aurora watching at very high latitudes was the main criticism of de Mairan's catalogue by Ramus,
 claiming that it relied on negative results from expeditions that were outside the observing season set by sunlight \citep{brekke94,stauning11}.

\subsubsection{Comparison of Aurorae during the Maunder and Dalton minima}
The debate about the reality of the drop in auroral occurrence during the Maunder minimum was ended when a decline
 was seen during the Dalton minimum (c. 1790-1830).
This minimum is seen in all the modern catalogues mentioned above and in others, such as that by \citet{nevanlinna95}
 from Finnish observatories, which can be calibrated against modern-day observations \citep{nevanlinna01}.
Many surveys show the MM to be deeper than the Dalton minimum in auroral occurrence but not by a very great factor \citep[e.g.,][]{silverman92}.
However, given the likelihood that aurorae were largely restricted to a narrow band at very high latitudes during both minima,
 observations at such high latitudes become vital in establishing the relative depths of these two minima.
In this respect the survey by \citet{vazquez14} is particularly valuable as, in addition to assigning locations to every sighting,
 it includes the high latitude catalogues by \citet{runebson1882} and \citet{tromholt1898} as well as those of \citet{silverman92}
  and \citet{fritz1873}.
The quality control employed by \citet{vazquez14} means that their survey extends back to only 1700 which implies that
  it covers 15 years before the events of 1716 and hence only the last solar cycle of the MM.

\begin{figure}[t]
\centering \resizebox{\columnwidth}{!}{\includegraphics{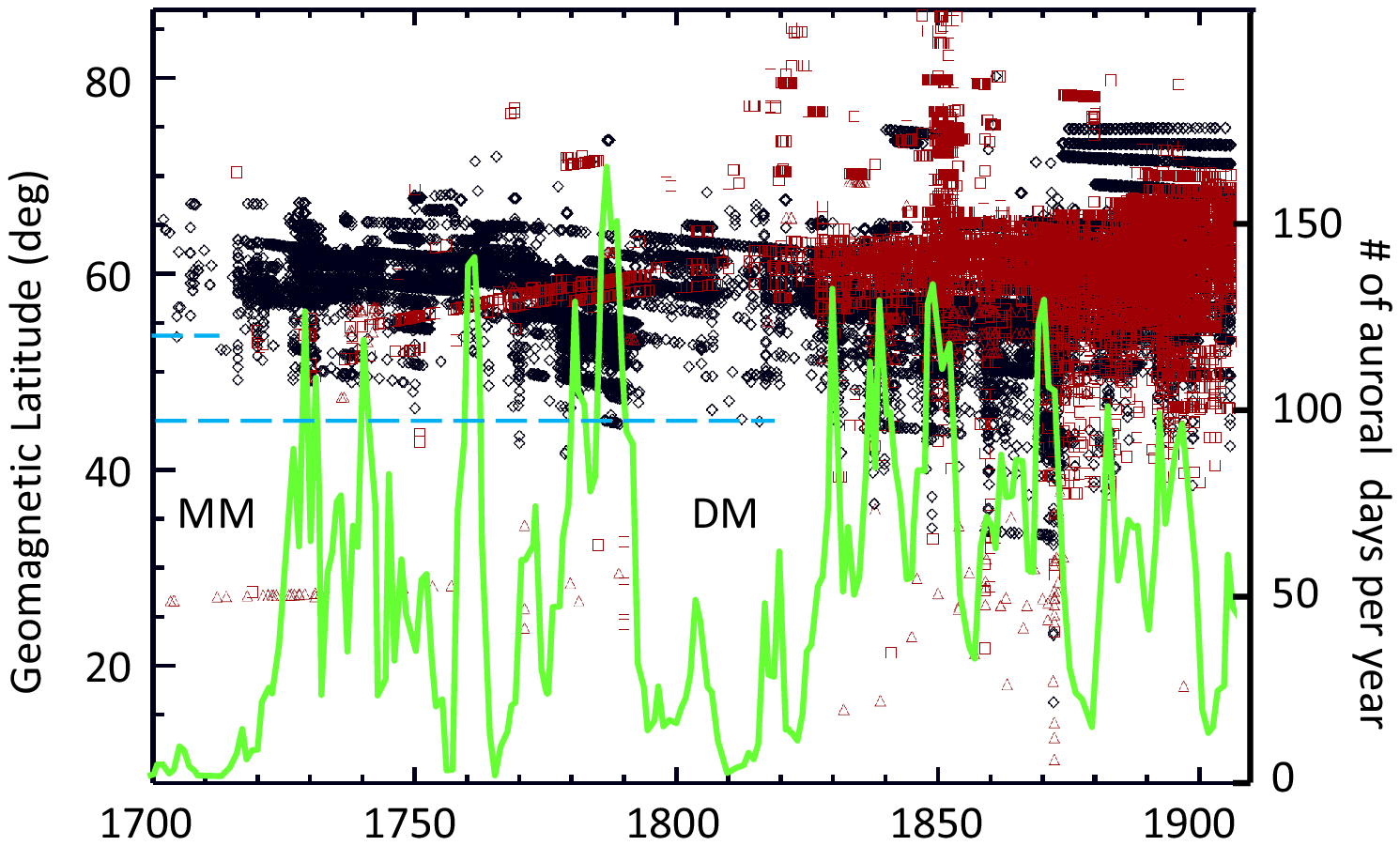}}
\caption{Occurrence of auroral reports, 1700-1900.
The green line is the number of auroral nights at geomagnetic latitudes below 56$^\circ$ from a combination of several catalogues
 \citep{nevanlinna95, fritz1873, fritz1881, legrand87}.
The points show the geomagnetic latitude and time of auroral sightings  from the catalogue of \citet{vazquez14} (their Figure 9).
Black diamonds, red squares and red triangles are, respectively, for observing sites in Europe and North Africa, North America, and Asia.
Blue dashed lines mark the minimum latitude of auroral reports in the last solar cycle of the Maunder minimum (MM) and in the two cycles of the Dalton Minimum (DM).
}
\label{Fig:aurorae}
\end{figure}
Figure~\ref{Fig:aurorae} is an analysis of the occurrence of aurorae between 1700 and 1900.
The green line is the number of auroral nights per year at geomagnetic
latitudes below 56$^\circ$ from a combination of the catalogues of \citet{nevanlinna95}, \citet{fritz1873},
 \citet{fritz1881} and \citet{legrand87}.
This sequence clearly shows that aurorae at geomagnetic latitudes below 56$^\circ$
 were indeed rarer in both the last cycle of the MM and the two cycles of the Dalton minimum (DM).
However, the number of recorded auroral sightings was significantly greater during DM than that in the MM.
The points in Figure~\ref{Fig:aurorae} show the geomagnetic latitude and time
 of auroral sightings from the catalogue of \citet{vazquez14} (their Figure 9).
Black diamonds, red squares and red triangles are, respectively, for observing
 sites in Europe and North Africa, North America, and Asia.
Blue dashed lines mark the minimum latitude of auroral reports in the last
 solar cycle of the MM and in the two cycles of the DM.
During the Dalton minimum many more aurorae were reported (symbols in the Figure) poleward of the
 56$^\circ$ latitude.

Considerably fewer arcs were reported at the end of the MM at these
 latitudes, despite the inclusion by \citet{vazquez14} of two extra catalogues of such
 events for this period at auroral oval latitudes.
Furthermore, the two dashed lines show the minimum latitude of events seen in
 these two minima: whereas events were recorded down to a magnetic latitude of 45$^\circ$ in the DM,
 none were seen at the end of the MM below 55$^\circ$, consistent with the dearth of observations in central Europe at the time.
Note that during the MM there are some observations at magnetic latitudes near
 27$^\circ$, all from Korea (with one exception which is from America).
They were reported to be observed in all directions, including the South, and to be red \citep{yau95} which makes
 them unfavorable candidates for classical aurorae.
By their features these could have been stable auroral red (SAR) arcs \citep{zhang85} which in modern times are
 seen at mid-latitudes mainly during the recovery phase of geomagnetic storms \citep{kozyra97}.
These arcs are mainly driven by the ring current and differ from the normal auroral phenomena.
Moreover, as stated by \citet{zhang85} ``We cannot rule out the possibility that some of these Korean sightings
 were airglows and the zodiacal light''.
We here concentrate on the higher latitude auroral oval arcs.
Note that the plot also shows the return of reliable lower latitude sightings in Europe in 1716 and in America in 1718.

\begin{figure}[t]
\centering \resizebox{\columnwidth}{!}{\includegraphics{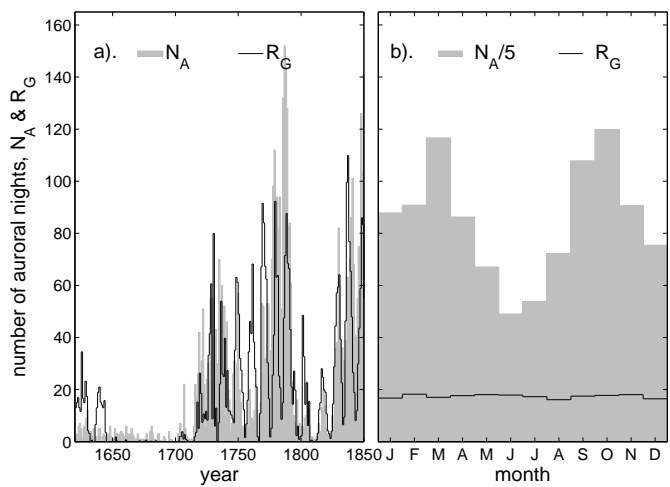}}
\caption{Same as Figure~\ref{Fig:lowe}, but compiled from 41 different catalogues
of auroral observations at magnetic latitudes below 55$^\circ$ in Europe, Asia,
 North Africa, New England and Great Britain.
The time series covers both the Maunder and the Dalton minima.
}
\label{Fig:all}
\end{figure}

Figure~\ref{Fig:all} corresponds to Figure~\ref{Fig:lowe}, but now based on a
compilation of all major historical auroral catalogues.
Figure~\ref{Fig:all} employs the list of aurora days by \citet{krivsky88} which is based on
39 different catalogues of observations at geomagnetic latitudes below 55$^\circ$
 in Europe, Asia, North Africa.
To this has been added the catalogue of \citet{lovering1867}\footnote{Paper and
catalogue are available from http://www.jstor.org/stable/25057995.} of observations made
 in and around Cambridge, Mass., USA which was at a magnetic latitude close to
 55$^\circ$ in 1900, and the recently-discovered catalogue of observations from
 Great Britain by \citet{lowe1870}.
Figure~\ref{Fig:all}a shows the low level and gradual decline in the occurrence
 of low- and mid-latitude auroral observations during the MM and in
 the decades leading up to it.
This is in contrast to the general rise in observation reports that exists on
longer timescales as scientific recording of natural phenomena became more common.
After the MM, solar cycles in the auroral occurrence can clearly be seen and
the correlation with the annual mean group sunspot numbers $R_G$ is clear.
Even for these lower latitude auroral observations it is unquestionable that the MM is
considerably deeper than the later Dalton minimum.
The annual variability (Figure~\ref{Fig:all}b) is obvious also for this data set.

From all of the above it is clear that the Maunder minimum in auroral, and hence
solar, activity was a considerably deeper minimum than the later Dalton minimum.

\subsection{Solar corona during the Maunder minimum}

As shown already by \citet{eddy76} and re-analyzed recently by \citet{riley15}, recorded observations
 of solar eclipses suggest the virtual absence of the bright structured solar corona during the MM.
While 63 total solar eclipses should had taken place on Earth between 1645 and 1715, only four
 (in years 1652, 1698, 1706 and 1708) were properly recorded in a scientific manner, others were either
 not visible in Europe or not described in sufficient detail.
These reports \citep[see details in][]{riley15} suggest that the solar corona was reddish and unstructured
 which was interpreted \citep{eddy76} as the F-corona (or zodiacal light) in the absence of the
 K-corona.
The normally structured corona reappeared again between 1708 and 1716, according to later solar eclipse
 observations, as discussed in \citet{riley15}.

Observations of the solar corona during total eclipses, although rare and not easy
 to interpret, suggest that the corona was very quiet and had shrunk during the MM, with no
 large scale structures such as streamers.
This also implies the decline of surface activity during the MM.

\subsection{Heliospheric conditions}
\label{S:phi}

The Sun was not completely quiet during the MM, and a certain level of heliospheric
activity was still present -- the heliosphere existed, the solar wind was blowing,
 the heliospheric magnetic field was there, although at a strongly reduced level
 \citep[e.g.][]{cliver98,caballero_10Be_04,mccracken14}.
Since heliospheric disturbances, particularly those leading to cosmic ray
modulation, are ultimately driven by solar surface magnetism
 \citep{potgieterLR}, which is also manifested through sunspot activity, cosmic
 ray variability is a good indicator of solar activity, especially
 on time scales longer than a solar cycle \citep{beer00,beer12,usoskin_LR_13}.
Here we estimate the heliospheric conditions evaluated for the period around the
MM, using different scenarios of solar activity, and compare those
 with directly measured data on cosmogenic isotopes in terrestrial and extra-terrestrial archives.

%\subsubsection{Open solar magnetic flux}

The open solar magnetic flux (OSF) is one of the main heliospheric parameters defining the heliospheric modulation of cosmic rays.
It is produced from surface magnetic fields expanding into the corona from
 where they are dragged out into the heliosphere by the solar wind.
Consequently, it can be modelled  using the surface distribution of sunspots and a model of the surface magnetic flux
  transport \citep{wang02}.
If their exact surface distribution is not known, the number of sunspots can also serve as a good input to OSF
 computations, using models of magnetic flux evolution \citep{solanki00,solanki02,lockwood_3_14}
 or with more complex surface flux transport simulations \citep[e.g.,][]{jiang11}.
Here we use the simpler, but nonetheless very successful model to calculate the OSF from the sunspot
 number series \citep{lockwood_3_14,lockwood_AG_14}.
This model quantifies the emergence of open flux from sunspot number using an analysis of the occurrence rate and magnetic flux
 content of coronal mass ejections as a function of sunspot number over recent solar cycles \citep{owens12a}.
The open flux fractional loss rate is varied over the solar cycle with the current sheet tilt, as predicted theoretically by \citet{owens11}
 and the start times of each solar cycle taken from sunspot numbers (during the MM the $^{10}$Be cycles are used).
The one free parameter needed to solve the continuity equation, and so model the OSF, is then obtained by
 fitting to the open flux reconstruction derived from geomagnetic data for 1845--2012 by \citet{lockwood_AG_14}.

\begin{figure}[t]
\centering \resizebox{\columnwidth}{!}{\includegraphics{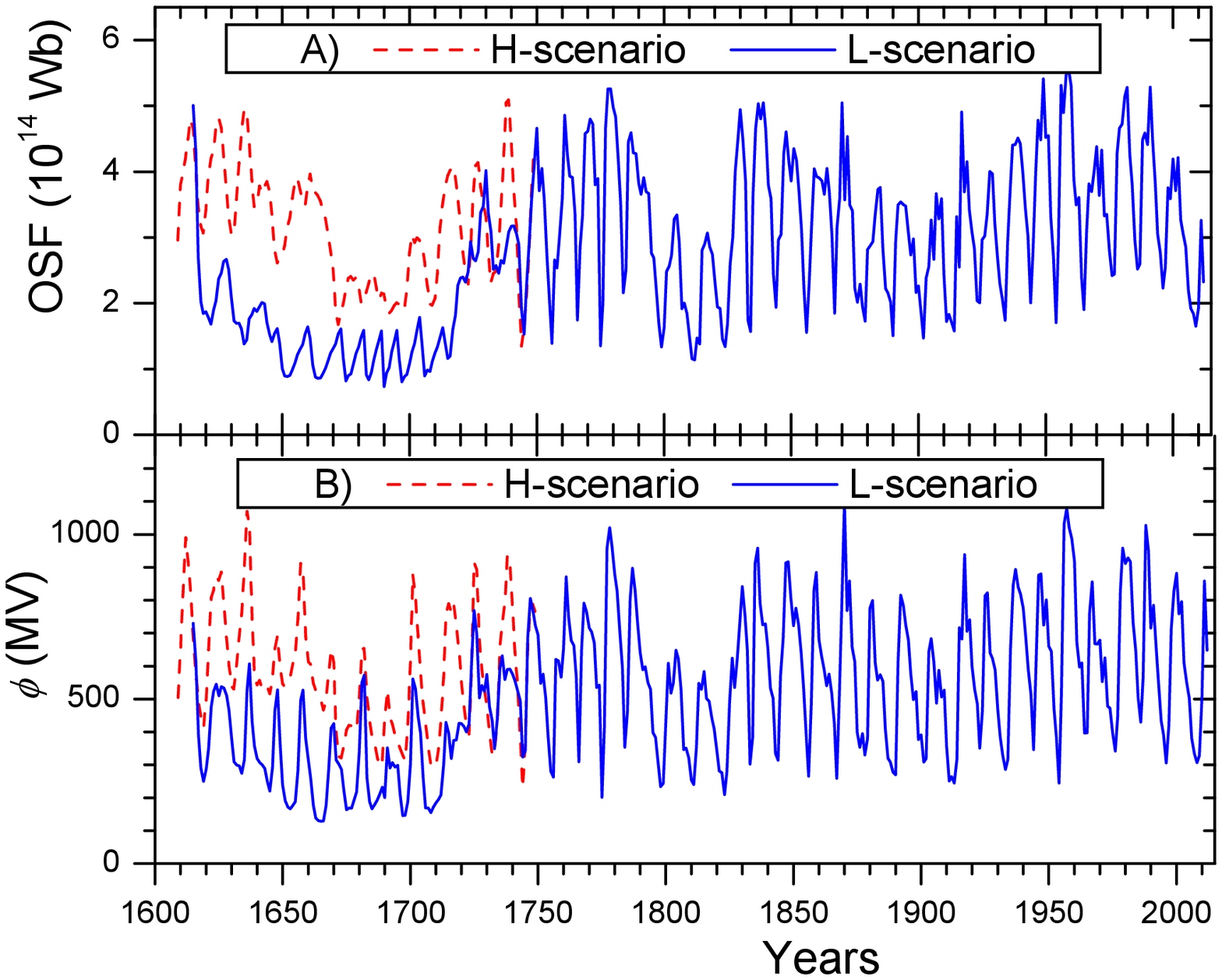}}
\caption{Evolution with time of the open solar magnetic flux, OSF, (panel A) and of the modulation potential
 $\phi$ (panel B) for the two scenarios of solar activity during the MM (see text).
}
\label{Fig:OSF}
\end{figure}

We computed OSF series (Figure~\ref{Fig:OSF}) corresponding to the two
scenarios for the number of sunspots during the MM (see Section~\ref{S:history}), viz. L- and H-scenarios.

The OSF is the main driver of the heliospheric modulation of cosmic rays on
 time-scales of decades to centuries \citep[e.g.,][]{usoskin_JGR_02},
 with additional variability defined by the heliospheric current sheet (HCS) tilt and the large scale polarity of
 the heliospheric magnetic field \citep{alanko06,thomas14}.
Using an updated semi-empirical model \citep{alanko06,asvestari15} we have computed the modulation potential $\phi$
 \citep[see definition and formalism in][]{usoskin_Phi_05} for the period since 1610 for the two scenarios described above,
 as shown in Figure~\ref{Fig:OSF}b.
These series will be used for a subsequent analysis.

\subsection{Cosmogenic radionuclides}
\label{S:cosmo}

The cosmogenic radionuclides are produced by cosmic rays in the atmosphere, and this forms the dominant source
 of these isotopes in the terrestrial system \citep{beer12}.
Production of the radionuclides is controlled by solar magnetic activity quantified in the heliospheric
 modulation potential (see above) and by the geomagnetic field,
 both affecting the flux of galactic cosmic rays impinging on Earth.
For independently known parameters of the geomagnetic field, one can use the measured abundance of cosmogenic radioisotopes in
 a datable archive to reconstruct, via proper modelling including production and transport of the isotopes in the Earth's
 atmosphere, the solar/heliospheric magnetic activity in the past \citep{beer12,usoskin_LR_13}.
Here we used a recent archeomagnetic reconstruction of the geomagnetic field \citep{licht13} before 1900.
In the subsequent subsections we apply the solar modulation potential series obtained for the two
 scenarios (Figure~\ref{Fig:OSF}b) to cosmogenic radionuclides.

\subsubsection{$^{14}$C in tree trunks}
\label{S:14C}
\begin{figure}[t]
\centering \resizebox{\columnwidth}{!}{\includegraphics{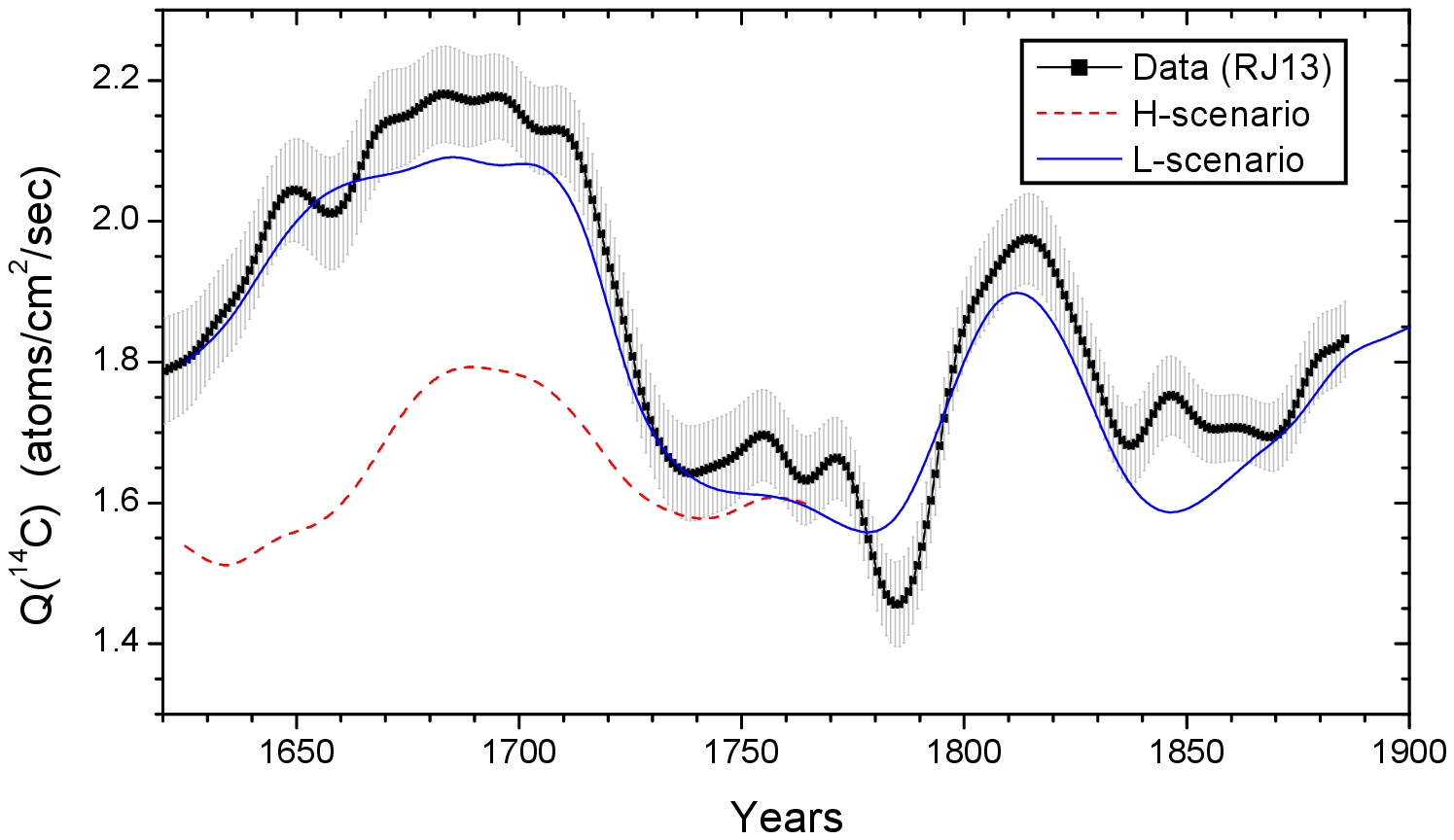}}
\caption{Time profile of decadally smoothed radiocarbon $^{14}$C production rate.
The black curve with grey error bars represents the reconstruction by \citet{roth13} based on the
 Intcal13 \citep{reimer13} radiocarbon data.
 Colored curves depict the computed production for the two scenarios as labeled in the legend (see text for details).
}
\label{Fig:14C}
\end{figure}

Using the recent model of radiocarbon $^{14}$C production \citep{kovaltsov12}, we have computed the expected global mean
 radiocarbon production rate for the two scenarios analyzed here, as shown in Fig.~\ref{Fig:14C}.
One can see that the variability of $^{14}$C production is quite different in the H-(red dotted) and L-(blue solid curve) scenarios.
While the former is rather constant, with only a weak maximum during the MM, even smaller than that for the
 Dalton minimum in the early 1800s, the latter exhibits a high and long increase during the MM, which is
 significantly greater than that for the DM in both amplitude and duration.

In the same plot we show also the $^{14}$C production rate obtained by \citet{roth13} from the Intcal13 \citep{reimer13}
 global radiocarbon data, using a new generation state-of-the-art carbon cycle model.
The $^{14}$C global production expected for the L-scenario agrees very well with the data, within the uncertainties,
 over the entire period of 1610--1880 confirming the validity of this scenario.
On the contrary, the H-scenario both quantitatively and qualitatively disagrees with the observed production during the MM, implying
 that the solar modulation of cosmic rays is grossly overestimated during the MM by this scenario.
Thus, the $^{14}$C data support a very low level of heliospheric (and hence solar surface magnetic) activity
 during the MM, a level that is considerably lower than during the Dalton minimum.

\subsubsection{$^{10}$Be in polar ice cores}

With a similar approach to that taken for the analysis of $^{14}$C
(section~\ref{S:14C}) we have computed the depositional flux of $^{10}$Be in polar regions.
We used the same archeomagnetic model \citep{licht13}, the recent $^{10}$Be production model by \cite{kovaltsov10}, and
 the atmospheric transport/deposition model as parameterized by \citet{heikkila09}.

\begin{figure}[t]
\centering \resizebox{\columnwidth}{!}{\includegraphics{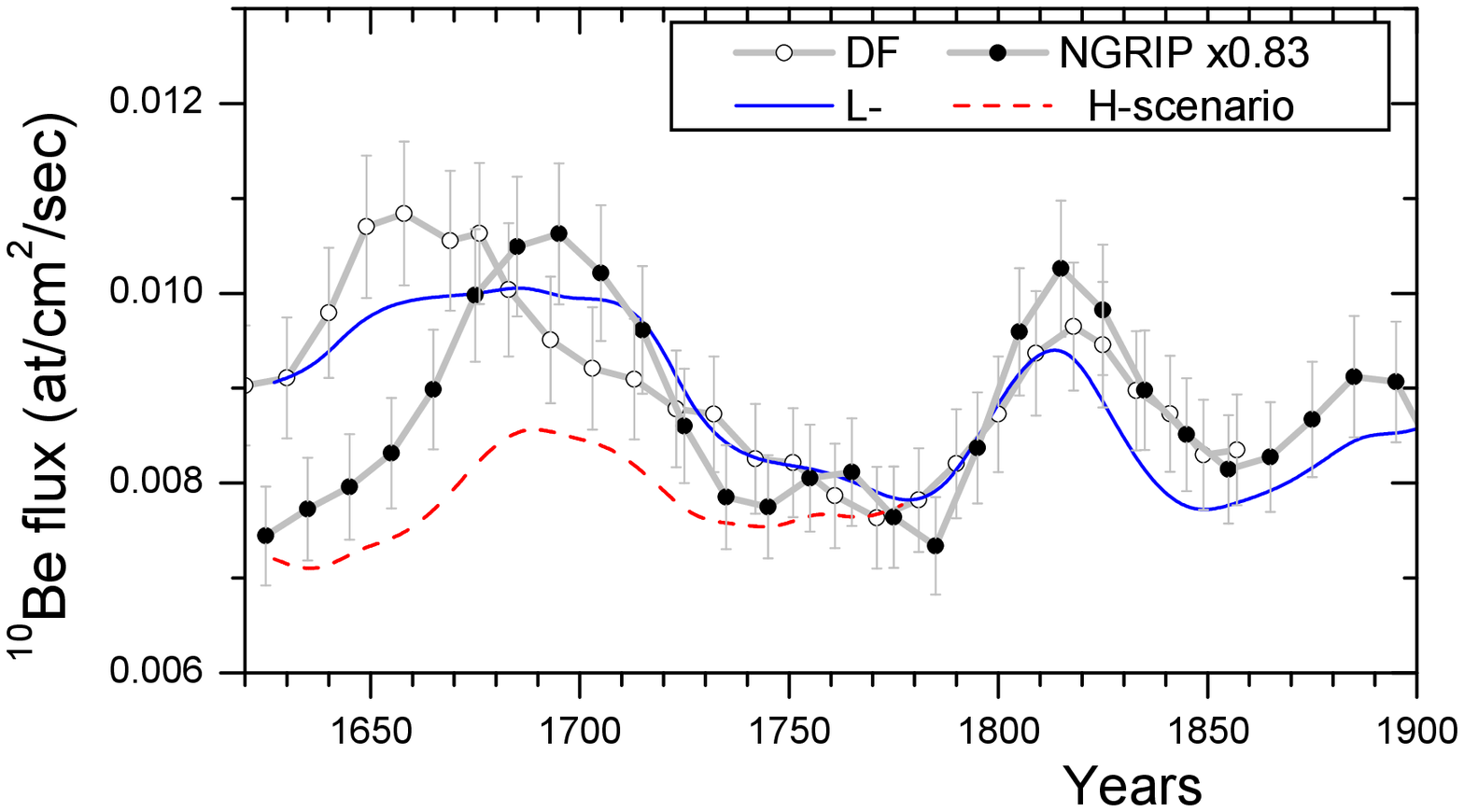}}
\caption{Depositional flux of $^{10}$Be in polar ice.
All data are pseudo-decadal and $\approx 25$-yr smoothed because of strong high-frequency noise.
Grey curves depict the fluxes measured at Dome Fuji (DF) in Antarctica \citep{horiuchi07}, and
 at NGRIP (scaled as 0.83), Greenland \citep{berggren09}.
Error bars are estimates of both statistical and systematic errors.
The blue and red curves depict the modelled $^{10}$Be depositional fluxes for the L- and H-scenarios (see text for details), respectively.
}
\label{Fig:10Be}
\end{figure}

The results are shown in Figure~\ref{Fig:10Be}.
As discussed in the previous section, the expected curve for the H-scenario (red dashed curve) shows little
variability, being lower (implying higher solar modulation) during the MM than during the DM, while the L-scenario
 yields a higher flux (lower modulation) during the MM.
The two grey curves depict $^{10}$Be fluxes measured in two opposite polar regions.
One is the data series of $^{10}$Be depositional flux measured in the Antarctic Dome Fuji (DF) ice core \citep{horiuchi07}.
The other is the $^{10}$Be flux series measured in the Greenland NGRIP (North Greenland Ice-core Project) ice core
 \citep{berggren09}.
Because of the different local climate conditions \citep{heikkila09}, the latter was scaled by a factor of 0.83 to match
 the same level.
This scaling does not affect the shape of the curve and in
 particular not the ratio of the $^{10}$Be flux in the MM and the DM.
One can see that while the time profiles of the two datasets differ in detail, probably because
 of the different climate patterns \citep{usoskin_10Be_09} and/or timing uncertainties, both yield high $^{10}$Be production during
 the MM.
This corresponds to extremely low solar activity \citep{mccracken04}.
The L-scenario agrees with the data reasonably well (the data display even higher
 maxima than the model), while the H-scenario clearly fails to reproduce the
 variability of $^{10}$Be measured in polar ice.

Thus, the $^{10}$Be data from both Antarctic and Greenland ice cores support
a very low level of heliospheric (and hence solar surface magnetic) activity
 during the MM, significantly lower than during the Dalton minimum.

\subsubsection{$^{44}$Ti in meteorites}

\begin{figure}[t]
\centering \resizebox{\columnwidth}{!}{\includegraphics{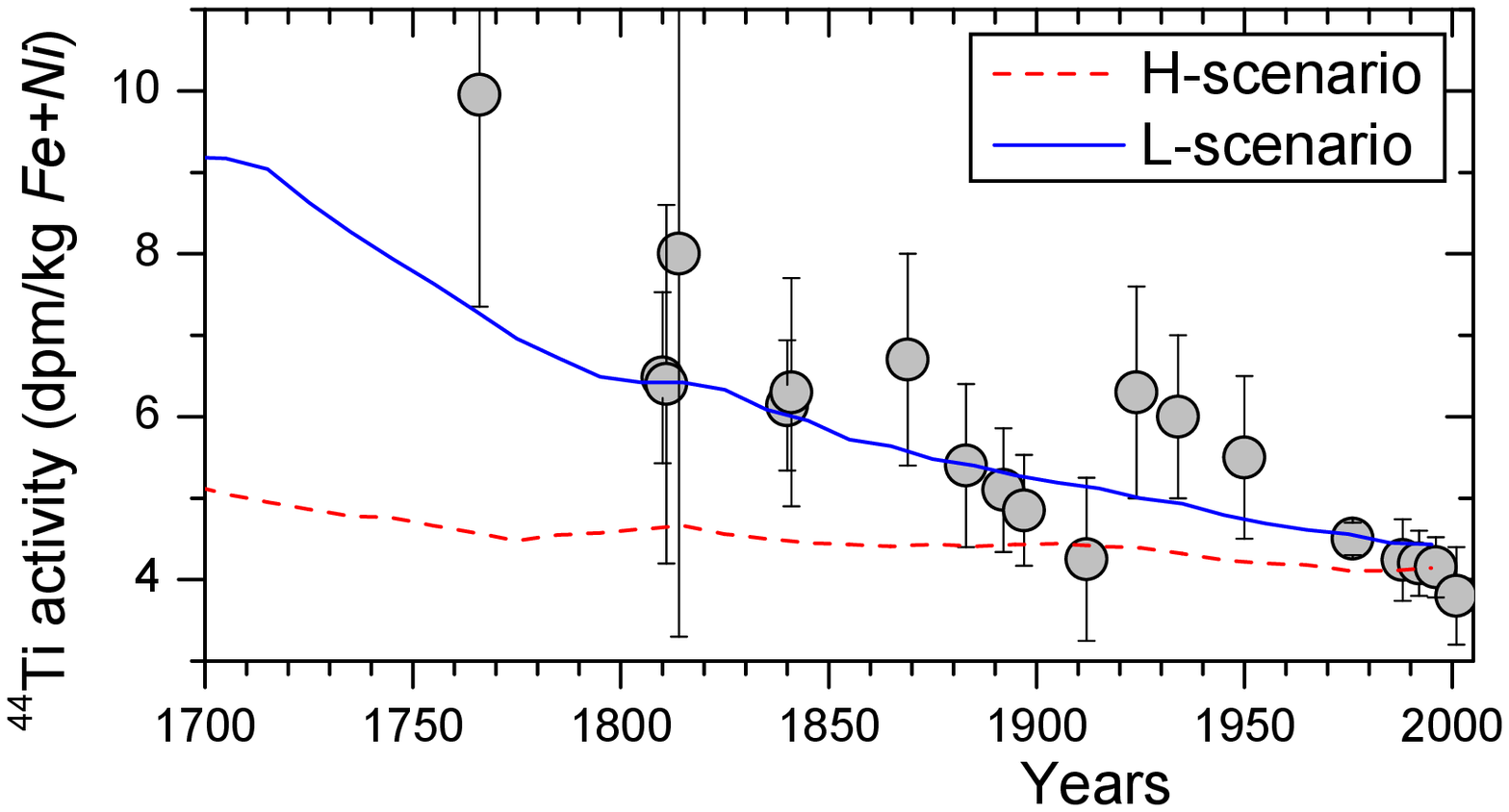}}
\caption{Time profile of the $^{44}$Ti activity in units of disintegrations per minute per
 kg of iron and nickel in the meteorite.
The grey dots with error bars are the measurements \citep{taricco06}.
The colored curves depict the computed activity for the two scenarios (see text).
}
\label{Fig:44Ti}
\end{figure}

While records of terrestrial cosmogenic radionuclides may be affected by transport and
 deposition processes, which are not always exactly known \citep{usoskin_10Be_09,beer12}, cosmogenic nuclides measured in
 fallen meteorites are free of this uncertainty, since the nuclides are produced directly in the meteorite's body while in space,
 and measured after their fall on the Earth.
However, time resolution is lost or at least greatly reduced in this
case, and the measured activity represents a balance between
 production and decay over the time before the fall of the meteorite.
An ideal cosmogenic nuclide for our purpose is $^{44}$Ti with a half-life of about 60 years \citep{ahmad98,bonino95}.
Here we test the cosmic ray variability as inferred from different scenarios of solar activity since 1600, following
 exactly the method described in detail in \citet{usoskin_Ti_06} and the dataset of $^{44}$Ti activity measured in
 19 stony meteorites fallen between 1776 and 2001 \citep{taricco06}.
Applying the modulation potential series as described in Sect.~\ref{S:phi} to the cosmic ray flux and calculating the expected
 $^{44}$Ti activity as a function of the time of the meteorite's fall, we compare the model computations for the
 two scenarios with the measured
 values in Fig.~\ref{Fig:44Ti}.
One can see that the series for the L-scenario fits the data rather well,
whereas the curve resulting from the H-scenario lies considerably too low.
As a merit parameter of the agreement we use the $\chi^2$ value for the period of 3.3 half-lives (attenuation factor 10)
 since the middle of the MM, viz. until 1880, which includes 7 meteorites fallen between 1776 and 1869 (7 degrees of freedom).
The $\chi^2(7)$ for the L-scenario is 2 (0.33 per degree of freedom), which perfectly fits the hypothesis.
For the H-scenario, however, $\chi^2(7)=17.6$ (2.93 per degree of freedom) indicating that this hypothesis should be rejected
 with a high significance of 0.014.
Note that the long residence times of the meteorites within the helisophere means that the difference during
 the MM between the two scenarios has an influence on the predicted $^{44}$Ti abundances even for meteorites that
 fell to Earth relatively recently.
This results in the observed abundances being inconsistent with the H-scenario for a large number of the meteorites
 whereas they are consistent, within the observational uncertainty, with the L-scenario.

Accordingly, the hypothesis of a high level of solar activity during the MM is rejected at a high significance level
 using indirect data of $^{44}$Ti in meteorites, while the conventional scenario of very low activity during the MM
 is in full agreement with the data.

%_________________________________
\section{Consequences of the Maunder minimum}
\label{S:cons}

\subsection{Solar/stellar dynamo}
\label{S:dynamo}
Major changes in the secular level of solar activity, such as Grand minima/maxima form a challenge for
 our understanding of the origin and evolution of the solar magnetic field.
It has been recently shown, by analyzing the sunspot numbers reconstructed from $^{14}$C for the last
 3000 years, that Grand minima form a separate
 mode of solar activity \citep{usoskin_AAL_14}, likely corresponding to a special regime of the solar dynamo.
In addition to the traditional concept of cyclic solar activity associated with periodic nonlinear oscillations of
 large-scale magnetic field, solar dynamo models include now, as their natural element, various deviations from
 pure periodicity.
Accordingly, it is crucially important for our understanding of solar/stellar dynamos to know whether Grand minima indeed
 exist and what their parameters are.
Although a possibility of direct modelling of the MM by dynamo theory is limited by the lack of information concerning
 flows in the solar interior, some important observational results pointing to quite peculiar features of the solar
 surface magnetic field configuration (slower, but more differential, rotation, strong hemispheric
 asymmetry of sunspot formation, and a possibly variable solar diameter) during the MM have been found \citep{ribes93,sokoloff94}.

Most large-scale solar dynamo models operate with averaged quantities taken as statistical ensembles of a moderate
 number of convective cells \citep[see, e.g., a review by][]{charbonneauLR} and include, in addition to the solar differential
 rotation, the collective inductive effect of mirror asymmetric convective turbulence and/or meridional circulation.
The observational knowledge of the turbulent quantities is especially limited, they have to be estimated
 using local direct numerical simulations \citep[e.g.][]{schrinner2005,KKB09}.
Generally, fluctuations of the intensity of the main drivers of the dynamo up to 10\%--20\% are expected.
By including such fluctuations into a mean-field dynamo models it is possible \citep[e.g.][]{moss08,choudhuri12,passos14a} to
 numerically reproduce a dynamo behavior which deviates from the stable cyclic evolution and
  depicts variability like the Grand minima.
It is important that asymmetric magnetic configurations of mixed parity with respect to the solar equator can be
 excited even in the framework of conventional $\alpha \Omega$ dynamo \citep{brandenburg89,jennings91},
  similar to that existing on the Sun at the end of the MM \citep{sokoloff94}.
The asymmetric sunspot occurrence during the late MM may be a signature of an unusual mode of the dynamo.
It is known for spherical dynamos \citep{moss_MNRAS_08} that even relatively moderate deviations from the nominal parameters
 associated with normal cycles can lead to the excitation of specific magnetic configurations, for example, with a
 quadrupolar symmetry with respect to the solar equator.
On the other hand, such an asymmetry is not expected for regular cycles  with normal values of the driving parameters.
This, along with the suppression of the cycle amplitude, may be a specific feature of a MM-type event.

An interesting fact is that the solar surface rotation was reported to be slower and more differential
 (changing faster with latitude) during the second half of the MM than during modern times \citep{ribes93}.
The enhanced differential rotation may have lasted until the mid-17th century \citep{arlt12}.
These facts are related to the operation of the solar dynamo during the MM.
Since the solar differential rotation is a main driver of the dynamo, and the asymmetry implies a specific configuration,
 the dynamo could had being operating in a special state during this period.

Differential rotation modulation and mixed parity are also an outcome of nonlinear dynamo models including
 Lorentz forces and the momentum equation.
In those cases, grand minima are produced without a stochastic dynamo effect \citep[][for spherical shells]{kueker99,pipin99,bushby06}.
The presence of grand minima is therefore a natural feature in mean-field dynamo
 modelling, including `side-effects' like differential rotation variation and mixed parity, and is a result
 of stochasticity and the nonlinearity of the full MHD equations.

Thus, the known phenomenology of the MM does not favor its interpretation as just a modulation of
 the normal 22-year Hale cycle by an additional longer cycle (Gleissberg cycle) as proposed, e.g., by ZP15.
The very asymmetric and suppressed sunspot activity, accompanied by slower differential rotation, during the late
 phase of the MM implies, in the light of dynamo theory, a special mode of dynamo operation, leading to Grand minima.

Meanwhile, it is still difficult to perform a direct numerical modelling of the MM, including sufficiently
 small scales to adequately model the convective turbulence \citep[e.g.][]{charbonneauLR}.
Although modern solar-type dynamo models reproduce the ``regular'' part of the behavior of the large-scale solar
 magnetic field \citep{ghizaru10,brown11,schrinner12,gastine12,KMB12}, it is still challenging
 to extend the integration time over several magnetic cycles as needed to reproduce a Maunder-like minimum
  \citep[see, however,][]{passos14,augustson15}.
Thus, while the MM is identified as a special mode of the solar dynamo, we are not yet able to precisely model it.

\subsection{Solar irradiance}
\label{S:irradiance}

\begin{figure}[th]
\centering \resizebox{\columnwidth}{!}{\includegraphics{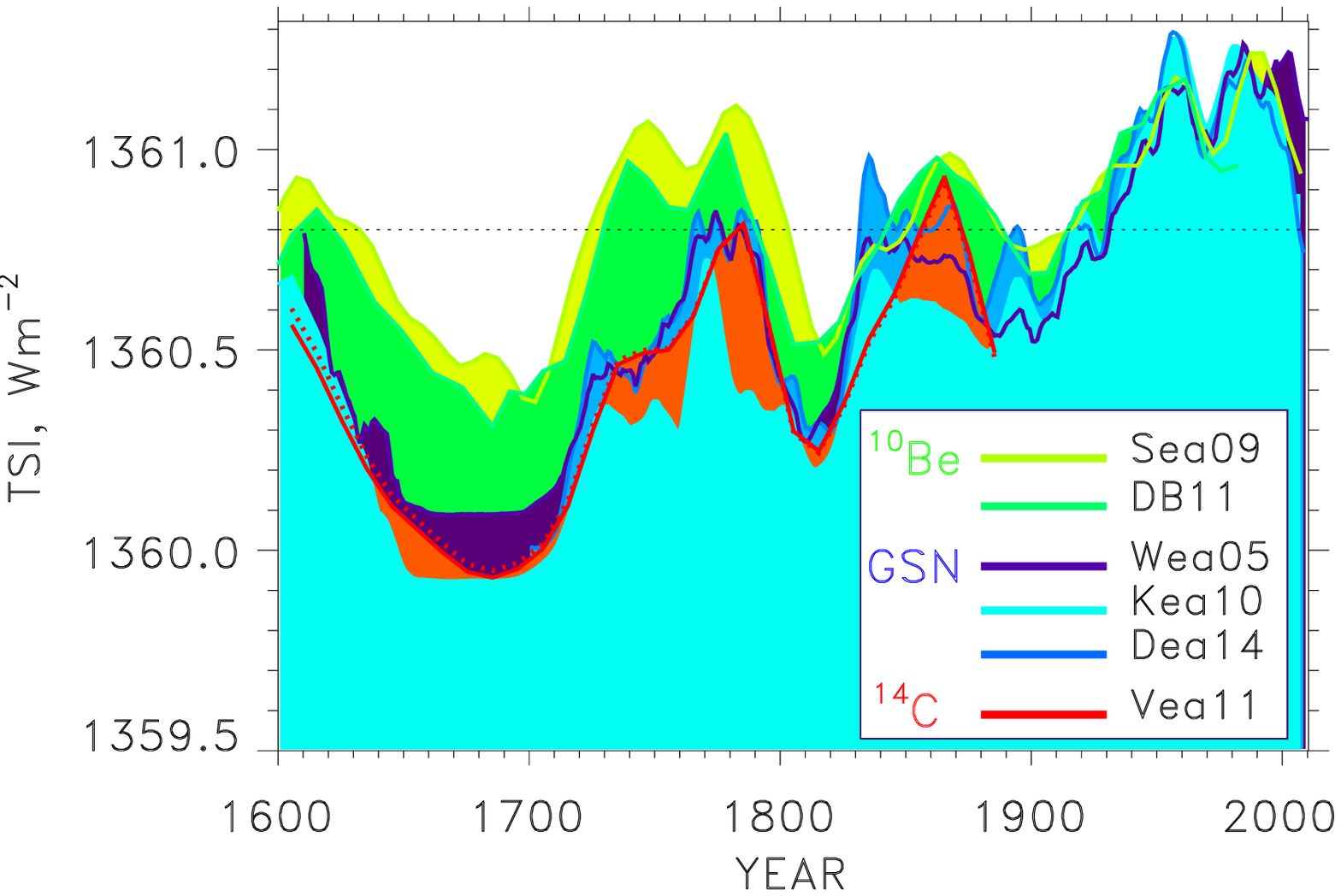}}
\caption{Selected TSI reconstructions since 1600, labeled in the plot are:
 Sea09 --\citet{steinhilber09}; DB11 -- \citet{delaygue11}; Wea05 -- \citet{wang05};
 Kea10 -- \citet{krivova10}; Dea14 -- \citet{dasi14}; Vea11 -- \citet{vieira11}.
The green, blue and red colour tones are used for the reconstructions based
on the $^{10}$Be, sunspot and $^{14}$C data, respectively.
The black dotted line marks the TSI value at modern solar activity minimum
conditions according to SORCE/TIM measurements.
}
\label{Fig:TSI_MM}
\end{figure}

It is now widely accepted that variations in solar irradiance in different wavelengths on time scales
 longer than about a day are driven by changes in the solar surface coverage
 by magnetic features, such as sunspots, that lead to a darkening of the
 solar disc, and faculae or network elements, that lead to a brightening,
 \citep[see, e.g.,][and references therein]{domingo09,solanki13}.
Consequently, the number of sunspots and faculae present during the MM would affect
 both the total (TSI) and the spectral (SSI) irradiance of the Sun at that special time in the history of solar activity.

Direct observations of TSI/SSI are available only from 1978 onwards \citep[e.g.,][]{froehlich13,kopp14}.
Consequently, a number of models have been developed that reconstruct solar irradiance back to the MM.
The TSI produced by a selection of such models is plotted in Fig.~\ref{Fig:TSI_MM}.
The selected models are based on very different data and techniques.

\citet{steinhilber09} and \citet{delaygue11} obtain their reconstructions from timeseries
 of $^{10}$Be concentrations in ice (green colour tones in the figure).
Both studies used a simple linear regression to calculate TSI.
\citet{steinhilber09} first estimated the interplanetary magnetic field from the $^{10}$Be
 data and then used a linear regression with the measured TSI following
 \citet{froehlich09}.
\citet{delaygue11} scaled the $^{10}$Be record by assuming a fixed change in TSI between the
 MM and the last decades.
Not shown in Fig.~\ref{Fig:TSI_MM} and also relying on the $^{10}$Be data is the TSI
 reconstruction by \citet{shapiro11}.
The magnitude of the secular change in this model comes from the difference between semi-empirical
 model atmospheres describing the darkest parts of the solar surface and the average quiet Sun
 (during modern times).
The secular change thus obtained is significantly higher than in all other models (3~W\,m$^{-2}$
 even after the re-assessment by \citealt{judge12}).
However, the shape of the secular change comes again from a linear regression to the $^{10}$Be record,
 so that it is essentially the same as in the other two models using $^{10}$Be concentrations.

Blue color tones show reconstructions that employ more physics-based approaches and are build on the sunspot number.
\citet{wang05} reconstructed the TSI from the magnetic flux evolved by a flux transport simulation \citep{sheeley05}.
\citet{krivova10} employed the approach of \citet{solanki02} to compute the magnetic flux from the sunspot number
 and therefrom the technique of \citet{krivova07} to compute the TSI and the spectral irradiance.
\citet{dasi14} used a flux transport model \citep{jiang10,jiang11} to simulate solar magnetograms
 to which they applied the SATIRE-S model \citep{fligge00,krivova03}.

Finally, \citet{vieira11} adapted the models by \citet{solanki02} and \citet{krivova07} for use
 with $^{14}$C data (red color tones).
The red lines show reconstructions by \citet{vieira11} based on two different models of the geomagnetic field.
They hardly diverge over the considered period.

The magnitude of the secular increase in the TSI since the MM differs by roughly 0.5~W\,m$^{-2}$
 between the models shown in the figure.
The full range of this variation is actually considerably larger, since the change obtained by
 \citealt{shapiro11}, in the correction proposed by \citealt{judge12}, is about 3~W\,m$^{-2}$.
More important for the purposes of the present paper is that despite this quantitative difference,
 the trends shown by all reconstructions are qualitatively similar.
In other words, quite irrespective of the data or the technique used for the TSI reconstruction, the TSI is
 always lower during the MM than during the Dalton minimum by on average 0.2--0.3~W\,m$^{-2}$
  (more for the model by \citealt{shapiro11}).
This appears to be a rather robust feature.
If the sunspot number during the MM were as high as proposed by ZP15 then this difference would
 vanish.
This contradicts all TSI reconstructions, including ones that are not built on sunspot numbers at
 all, although the uncertainties in the TSI reconstructions are sufficiently large that this cannot
 be judged to be a very stong constraint (and is not entirely independent of the other arguments
 provided in this paper).

%_________________________________
\section{Conclusions}
\label{S:discuss}

We have revisited the level of solar activity during the Maunder minimum, using all the existing, both direct and indirect,
 datasets and evidence to show that the activity was very low, significantly lower than during
 the Dalton minimum or the current weak solar cycle \# 24.
We have confronted the data available with two scenarios of the solar activity level during the MM -- the
 low activity (L-scenario) and the high activity (H-scenario, see Sect.~\ref{S:history}).

The results can be summarized as follows:

We have evaluated (Sect.~\ref{S:FA}), using a conservative approach, the fraction of sunspot active days during the MM.
The fraction appeared small, implying a very low level of sunspot activity.

We have revisited (Sect.~\ref{S:telescop}) the telescopic solar observations during the MM and we conclude that the
 astronomers of the 17th century, especially in its second half, were very unlikely to be influenced by the religious
 or philosophical dogmas.
This is contrary to the claim by ZP15.

We have discussed that the short gaps in the HS98 database, which were interpreted by ZP15 as deliberate omission of sunspot records
 by the 17th century observers for non-scientific reasons, are caused by a technical artefact of the database compilation and do
 not correspond to observational lacunas.

We have pointed out (Sect.~\ref{S:error}) outdated and erroneous information and serious methodological flaws in the analysis done by ZP15
 that led them to severely overestimate the solar activity level during the MM.

The latitudinal extent of sunspot formation (Sect.~\ref{S:lat}) also points to a very low solar activity during the MM.

We have shown (Sect.~\ref{S:naked}) that East-Asian naked-eye sunspot observations cannot be used to assess the exact
 level of solar activity.
The existing data do not contradict to the very low activity level during the MM.

We have presented a re-analysis of several documented sources, some new to science so far, on the occurrence of lower-latitude aurorae on Earth.
We have demonstrated that the MM indeed displayed very low activity also in terms of auroral sightings (Sect.~\ref{S:aurorae})
 compared to normal periods and also to the Dalton minimum.

We have compared (Sect.~\ref{S:cosmo}) the estimated heliospheric conditions (Sect.~\ref{S:phi}) for the two scenarios,
 with the actually measured cosmogenic isotope data for the period around the MM.
The comparison is fully consistent with the L-scenario but rejects the H-scenario at a very high confidence level.

We have argued (Sect.~\ref{S:dynamo}) that the observational facts (very low sunspot activity, hemispheric asymmetry of sunspot formation,
 unusual differential rotation of the solar surface and the lack of the K-corona) imply a special
 mode of the solar dynamo during the MM, and disfavor an interpretation of the latter as a regular minimum of the
 centennial Gleissberg cycle as proposed by ZP15.

We have discussed (Sect.~\ref{S:irradiance}) consequences of the MM for the solar irradiance variability which is a
 crucial point for the assessment of solar variability influence on both global and regional climate \citep{lockwood12,solanki13}.

We conclude, after careful revision of all the presently available datasets for the Maunder minimum, that
 solar activity was indeed at an exceptionally low level during that period, corresponding to a special
 Grand minimum mode of solar dynamo.
The suggestions of a moderate-to-high level of solar activity during the Maunder minimum
 are rejected at a high significance level.

\acknowledgement{
We are grateful to Regina v. Berlepsch, Potsdam, and Pierre Leich, N\"urnberg,
 the ECHO digital heritage online, and the Swiss Electronic Library, E-lib.ch, for providing electronic versions of a number of historic
 publications and manuscripts.
We are very grateul to Ralph Neuh\"auser for pointing to the original work by Marius, to Clara Ricken and Yori Fournier for their help with translations of original sunspot records.
This work was partly done in the framework of the ReSoLVE Centre of Excellence (Academy of Finland, project no. 272157) and partly
 supported by the BK21 plus program through the National Research Foundation (NRF) funded by the Ministry of Education of Korea.
J.M.V. acknowledges the support from the Junta de Extremadura (Research Group Grants GR10131) and from the Spanish Government (AYA2011-25945 and AYA2014-57556-P).
D.S. acknowledges Project RFFI 15-02-01407.
W.S. received no funding for his contribution to this paper.
Support from the COST Action ES1005 TOSCA (www.tosca-cost.eu) is gratefully acknowledged.
}

%++++++++++++++++++++++++++++++++++++++++++++++++++++++
%\bibliographystyle{apj}
%\bibliographystyle{aa} % style aa.bst
%\bibliography{J:/USOSKIN/papers/usoskin_all}
%\bibliography{C:/DATA_/USOSKIN/papers/usoskin_all}

\end{document}